\pdfoutput=1

\documentclass[10pt,twocolumn,letterpaper]{article}

\usepackage[pagenumbers]{cvpr} 

\usepackage[dvipsnames]{xcolor}

\usepackage[table]{colortbl}
\usepackage{verbatim}
\usepackage{graphicx}
\usepackage{setspace}

\usepackage{enumitem}

\definecolor{cvprblue}{rgb}{0.21,0.49,0.74}
\usepackage[pagebackref,breaklinks,colorlinks,citecolor=cvprblue]{hyperref}

\def\paperID{13807} 
\def\confName{CVPR}
\def\confYear{2024}

\title{SuGaR: Surface-Aligned Gaussian Splatting for\\ Efficient 3D Mesh Reconstruction and High-Quality Mesh Rendering}

\author{
Antoine Guédon \qquad Vincent Lepetit\\
LIGM, Ecole des Ponts, Univ Gustave Eiffel, CNRS, France\\
{\tt\small \url{https://anttwo.github.io/sugar/}}
}

\usepackage{dsfont}
\usepackage{etoolbox}
\usepackage{color}

\newif\ifshowedits

\newcommand{\addeditor}[3]{%
  \definecolor{#1color}{rgb}{#3}
  \expandafter\newcommand\csname #1\endcsname[1]{%
  \ifshowedits
    {\color{#1color} ##1}%
  \else
    {##1}%
  \fi
  }%
  \expandafter\newcommand\csname #1rmk\endcsname[1]{%
  \ifshowedits
    {\color{#1color} {\bf [#2: ##1]}}
  \fi
  }%
  \expandafter\newcommand\csname #1rpl\endcsname[2]{%
  \ifshowedits
    {\color{#1color} ##1 \sout{##2}}
  \else
    {##1}
  \fi
  }%
}

\newcommand{\createtextvar}[1]{
  \expandafter\newcommand\csname #1\endcsname{%
  {\text{#1}}
}%
}
\newcommand{\mycomment}[1]{}

\newcommand{\calN}{{\cal N}}

\newcommand{\calP}{{\cal P}}

\newcommand{\calR}{{\cal R}}

\newcommand{\IR}{{\mathds{R}}}

\newcommand{\vcomment}[1]{}

\addeditor{vincent}{VL}{0,0.5,0}
\addeditor{antoine}{AG}{0, 0, 0.8}

\begin{document}


\twocolumn[{%
\renewcommand\twocolumn[1][]{#1}%
\maketitle
\vspace{-1cm}
\begin{center}
    \captionsetup{type=figure}
\label{fig:teaser_two}
\begin{tabular}{@{}ccc@{}}
    \includegraphics[height=0.255\linewidth]{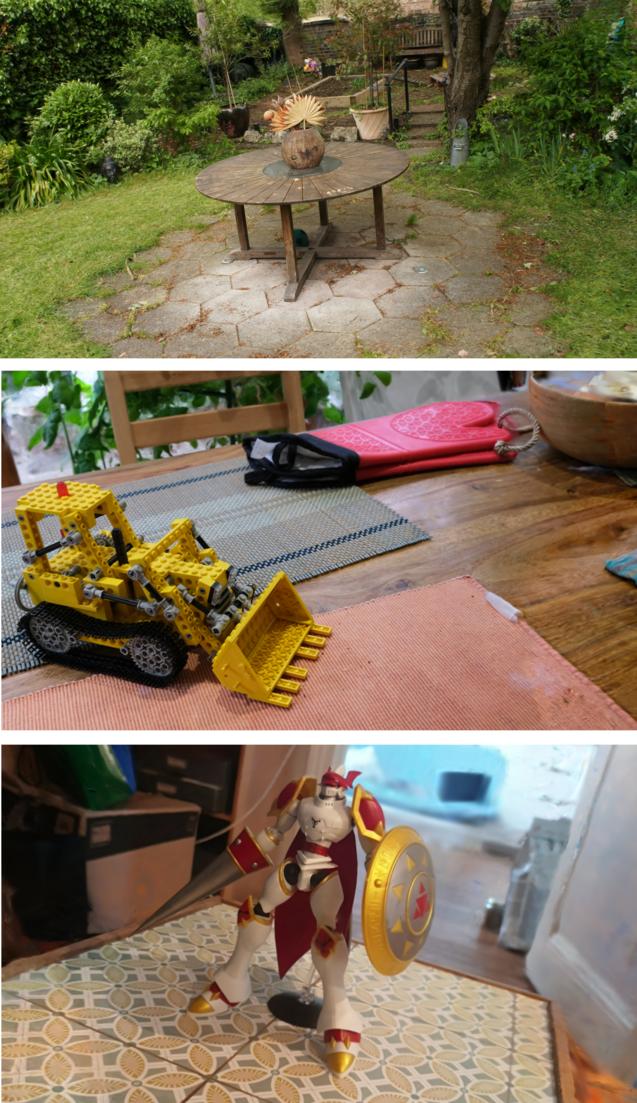} &
    \includegraphics[height=0.255\linewidth]{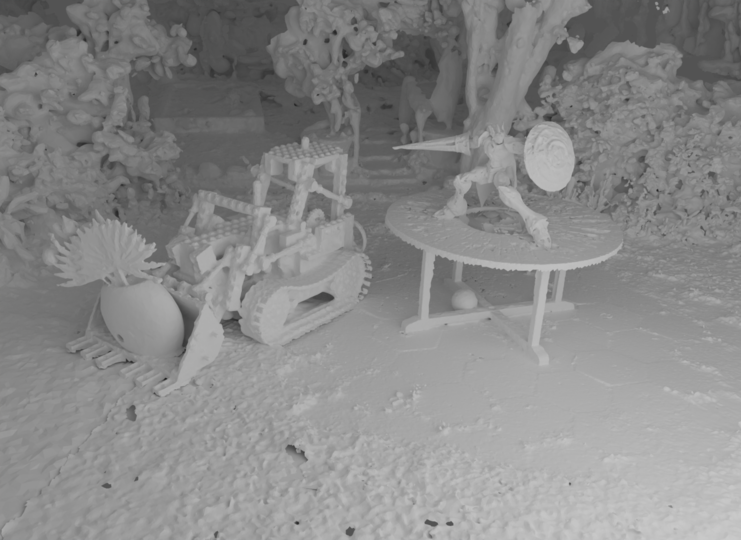} &
    \includegraphics[height=0.255\linewidth]{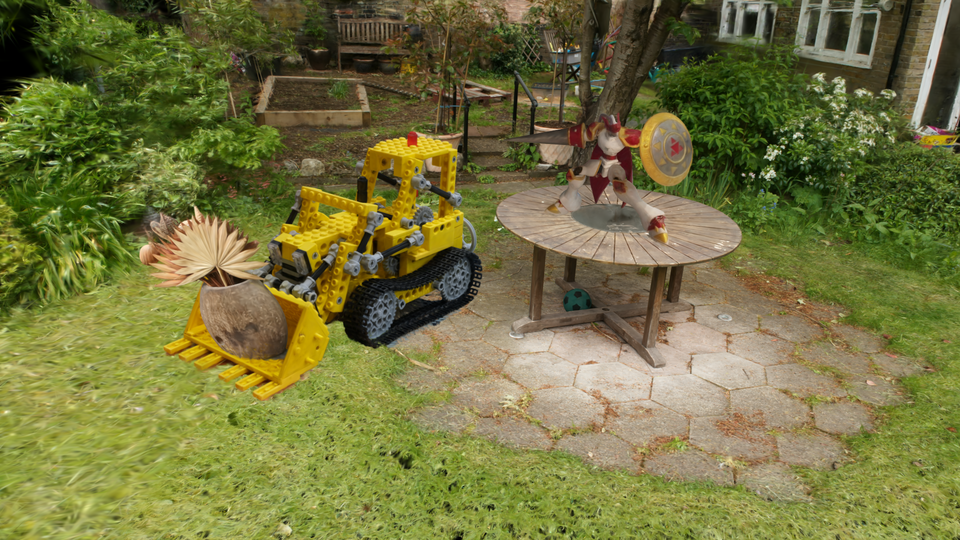} \\
\end{tabular}
\end{center}
\vspace{-0.6cm}
    \captionof{figure}{\label{fig:teaser_editing} We introduce a method that extracts accurate and editable meshes from 3D Gaussian Splatting representations within minutes on a single GPU.  The meshes can be edited, animated, composited, etc. with very realistic Gaussian Splatting rendering, offering new possibilities for Computer Graphics. Note for example that we changed the posture of the robot between the captured scene on the bottom left and the composited scene on the right.  The supplementary material provides more examples, including a video illustrating our results.}
\vspace*{0.5cm}
}]

{
\setstretch{0.97} 

\begin{abstract}
\vspace*{-0.5cm}
We propose a method to allow precise and extremely fast mesh extraction from 3D Gaussian Splatting~\cite{kerbl3Dgaussians}. Gaussian Splatting has  recently become very popular as it yields realistic rendering while being significantly faster to train than NeRFs. It is however challenging to extract a mesh from the millions of tiny 3D Gaussians as these Gaussians tend to be unorganized after optimization and no method has been proposed so far. Our first key contribution is a regularization term  that encourages the Gaussians to align well with the surface of the scene. We then introduce a method that exploits this alignment to extract a mesh from the Gaussians using Poisson reconstruction, which is fast, scalable, and preserves details, in contrast to the Marching Cubes algorithm usually applied to extract meshes from Neural SDFs. Finally, we introduce an optional refinement strategy that binds Gaussians to the surface of the mesh, and jointly optimizes these Gaussians and the mesh through Gaussian splatting rendering. This enables easy editing, sculpting, animating, and relighting of the Gaussians by manipulating the mesh instead of the Gaussians themselves. Retrieving such an editable mesh for realistic rendering is done within minutes with our method, compared to hours with the state-of-the-art method on SDFs, while providing a better rendering quality.
\end{abstract}

\begin{figure*}
\begin{center}
        \begin{tabular}{cc}
        \includegraphics[width=0.47\linewidth]{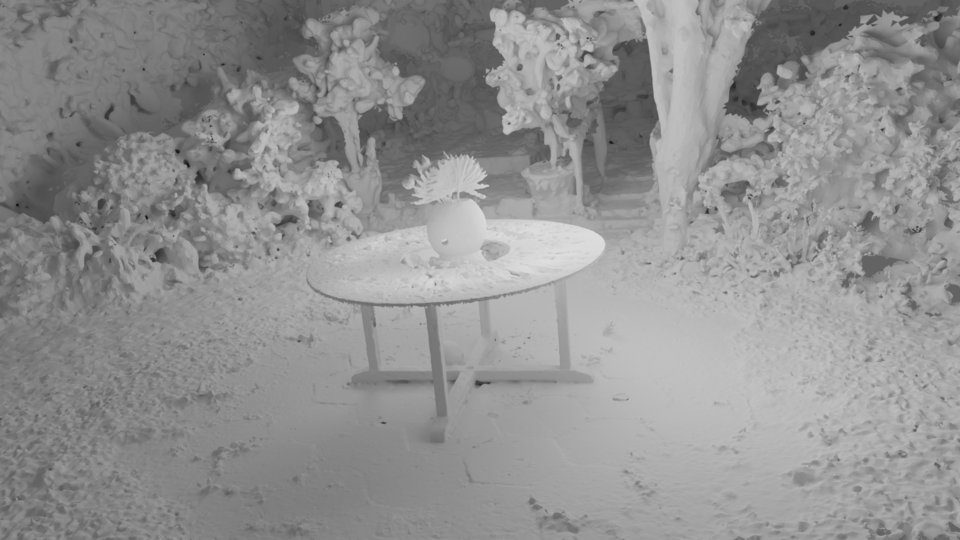} &
        \includegraphics[width=0.47\linewidth]{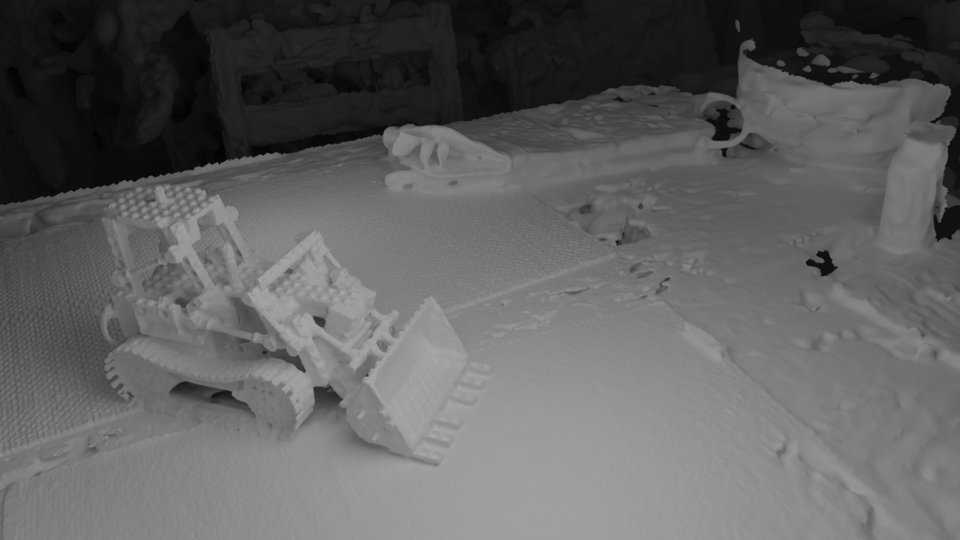} \\
        \includegraphics[width=0.47\linewidth]{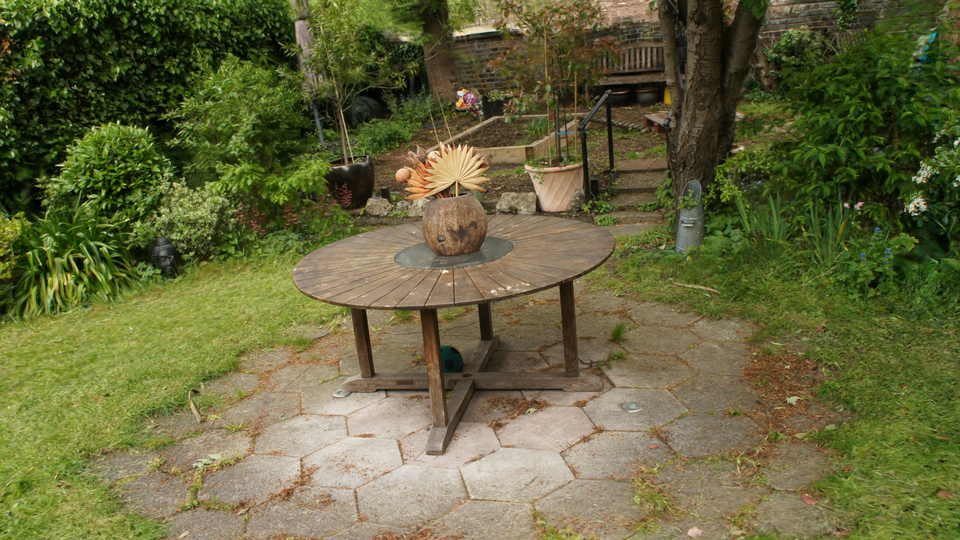} &
        \includegraphics[width=0.47\linewidth]{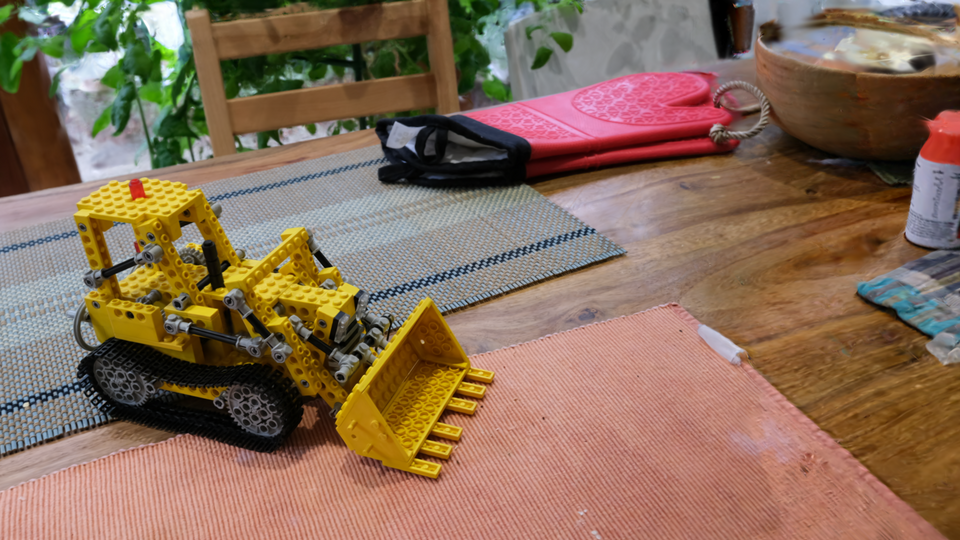}\\
    \end{tabular}
\end{center}
\vspace{-0.5cm}
\caption{
Our algorithm can extract a highly detailed mesh from any 3D Gaussian Splatting scene~\cite{kerbl3Dgaussians} within minutes on a single GPU~(\textbf{top:} Renderings of our meshes without texture, \textbf{bottom:} Renderings of the meshes with bound Gaussians).}
\label{fig:teaser_mesh}
\end{figure*}

\section*{Erratum}
\label{sec:erratum}
We identified a minor typographical error in Subsection~\ref{subsec:density_definition} in the earlier version of the paper.\\
In the computation of our regularization term $\calR$ in Equation~\ref{eq:R}, we use $p \rightarrow \pm s_{g*} \sqrt{-2 \log \left( d(p) \right)}$ instead of
$p \rightarrow \pm s_{g*} \sqrt{-2 \log \left( \bar{d}(p) \right)}$ as an 'ideal' distance function associated with the density $d$ (Equation~\ref{eq:gaussian_splatting_sdf_general}). 
As detailed in the paper, this distance function aligns with the true surface of the scene in an ideal scenario where $d = \bar{d}$.
We have updated Equation~\ref{eq:gaussian_splatting_sdf_general} to clarify this matter.

\section{Introduction}
\label{sec:intro}

\vspace*{0.2cm} 

After NeRFs~\cite{mildenhall2020nerf}, 3D Gaussian Splatting~\cite{kerbl3Dgaussians} has recently become very popular for capturing a 3D scene and rendering it from novel points of view. 3D Gaussian Splatting optimizes the positions, orientations, appearances (represented as spherical harmonics), and alpha blending of many tiny 3D Gaussians on the basis of a set of training images of the scene to capture the scene geometry and appearance. Because rendering the Gaussians is much faster than rendering a neural field, 3D Gaussian Splatting is much faster than NeRFs and can capture a scene in a few minutes.

\vspace*{0.1cm} 

While the Gaussians allow very realistic renderings of the scene, it is still however challenging to extract the surface of the scene from them: As shown in Figure~\ref{fig:one}, after optimization by 3D Gaussian Splatting, the Gaussians do not take an ordered structure in general and do not correspond well to the actual surface of the scene. In addition to the surface itself, it is also often desirable to represent the scene as a mesh, which remains the representation of choice in many pipelines: A mesh-based representation  allows for powerful tools for editing, sculpting, animating, and relighting the scene. Because the Gaussians after Gaussian Splatting are unstructured, it is very challenging to extract a mesh from them.  Note that this is also challenging with NeRFs albeit for different reasons.

In this paper, we first propose a regularization term that encourages the Gaussians to be well distributed over the scene surface so that the Gaussians capture much better the scene geometry, as shown in Figure~\ref{fig:one}. Our approach is to derive a volume density from the Gaussians under the assumption that the Gaussians are flat and well distributed over the scene surface. By minimizing the difference between this density and the actual one computed from the Gaussians during optimization, we encourage the 3D Gaussians to represent well the surface geometry.

Thanks to this regularization term, it becomes easier to extract a mesh from the Gaussians. In fact, since we introduce a density function to evaluate our regularization term, a natural approach would be to extract level sets of this density function. However, Gaussian Splatting performs densification in order to capture details of the scene with high fidelity, which results in a drastic increase in the number of Gaussians. Real scenes typically end up with one or several millions of 3D Gaussians with different scales and rotations, the majority of them being extremely small in order to reproduce texture and details in the scene. 
This results in a density function that is close to zero almost everywhere, and the Marching Cubes algorithm~\cite{lorensen-1987-marchingcubes}
fails to extract proper level sets of such a sparse density function even with a fine voxel grid, as also shown in Figure~\ref{fig:one}.

Instead, we introduce a method that  very efficiently samples points on the visible part of a level set of the density function, allowing us to run the Poisson reconstruction algorithm~\cite{kazhdan-2006-poissonsurfacereconstruction} on these points to obtain a triangle mesh. This approach is scalable, by contrast with the Marching Cubes algorithm for example, and reconstructs a surface mesh within minutes on a single GPU, compared to other state of the art methods relying on Neural SDFs for extracting meshes from radiance fields, that require at least 24 hours on one GPU~\cite{wang2021neus, yariv2021volsdf, li-cvpr2023-neuralangelo, yariv-2023-bakedsdf} and rely on multiple GPUs to speed up the process~\cite{rakotosaona2023nerfmeshing}.

\begin{figure}
    \centering
{\footnotesize
\begin{tabular}{c@{$\;\;$}c@{$\;$}c}
    \multicolumn{3}{c}{without our regularization term}\\
    \includegraphics[trim={7.33cm 0 8.5cm 0},clip,width=0.3\linewidth]
                    {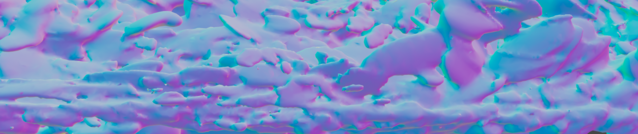} &
    \includegraphics[trim={0 0 0 0},clip,width=0.3\linewidth]
                    {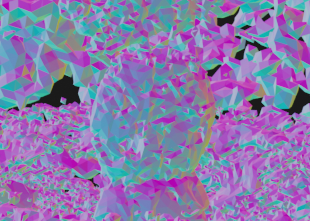} &
    \includegraphics[trim={0 0 0 0},clip,width=0.3\linewidth]
                    {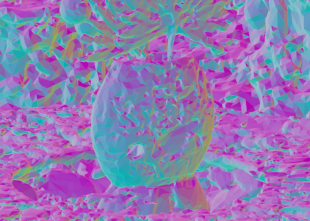}
                    \\
\hline                    
    \multicolumn{3}{c}{with our regularization term}\\
    \includegraphics[trim={7.33cm 0 8.5cm 0},clip,width=0.3\linewidth]
                    {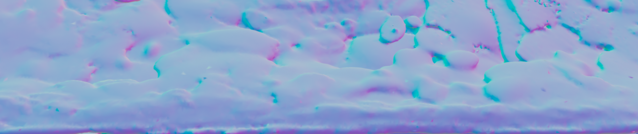} &
    \includegraphics[trim={0 0 0 0},clip,width=0.3\linewidth]
                    {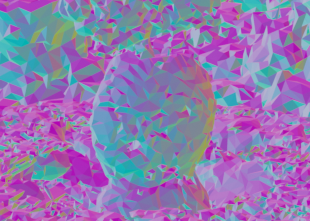} &
    \includegraphics[trim={0 0 0 0},clip,width=0.3\linewidth]
                    {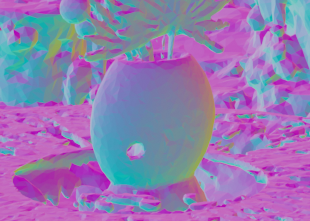} \\
 zoom on Gaussians & mesh with & mesh with our\\[-1mm]
 on a planar surface& Marching Cubes & extraction method\\                    
\end{tabular}
}
    \caption{\textbf{Extracting a mesh from Gaussians.}  Without regularization, the Gaussians have no special arrangement after optimization, which makes extracting a mesh very difficult.  Without our regularization term, Marching Cubes fail to extract an acceptable mesh. With our regularization term, Marching Cubes recover an extremely noisy mesh even with a very fine 3D grid.  Our scalable extraction method obtains a mesh even without our regularization term. Still, the mesh is noisy. By contrast, our full method succeeds in reconstructing an accurate mesh very efficiently.}
    \label{fig:one}
\end{figure}

As illustrated in Figures~\ref{fig:teaser_mesh} and~\ref{fig:sugar_renders}, our method produces high quality meshes. The challenge is in efficiently identifying points lying on the level set. To do this, we rely on the Gaussians depth maps seen from the training viewpoints. These depth maps can be obtained by extending the Gaussian Splatting rasterizer, and we show how to accurately sample points on the level set starting from these depth maps.  

Finally, after extracting this mesh, we propose an optional refinement strategy that jointly optimizes the mesh and a set of 3D Gaussians through Gaussian splatting rendering only. This optimization enables high-quality rendering of the mesh using Gaussian splatting rendering rather than traditional textured mesh rendering. This results in higher performance in terms of rendering quality than other radiance field models relying on an underlying mesh at inference~\cite{yariv-2023-bakedsdf, chen2022mobilenerf, rakotosaona2023nerfmeshing}. As shown in Figure~\ref{fig:teaser_editing}, this makes possible the use of traditional mesh-editing tools for editing a Gaussian Splatting representation of a scene, offering endless possibilities for Computer Graphics.

To summarize, our contributions are:
\begin{itemize}[noitemsep,topsep=0pt,parsep=0pt,partopsep=0pt,leftmargin=*]
    \item a regularization term that makes the Gaussians capture accurately the geometry of the scene;
    \item an efficient algorithm that extracts an accurate mesh from the Gaussians within minutes;
    \item a method to bind the Gaussians to the mesh, resulting in a more accurate mesh, higher rendering quality than state of the art methods using a mesh for Novel View Synthesis~\cite{yariv-2023-bakedsdf, rakotosaona2023nerfmeshing, chen2022mobilenerf}, and allowing editing the scene in many different ways.
\end{itemize}

We call our approach SuGaR.  In the remainder of the paper, we discuss related work, give a brief overview of vanilla 3D Gaussian Splatting, describe SuGaR, and compare it to the state of the art.

}

{
\setstretch{0.98} 
\section{Related Work}
\label{sec:relatedworks}

Image-based rendering~(IBR) methods rely on a set of two-dimensional images of a scene to generate a 
representation of the scene and render novel views. The very first novel-view synthesis approaches were based on light fields~\cite{levoy1996light}, and developed the concept of volume rendering for novel views. Their work emphasized the importance of efficiently traversing volumetric data to produce realistic images.

Various scene representations have been proposed since, such as triangle meshes, point clouds, voxel grids, multiplane images, or neural implicit functions.

\paragraph{Traditional mesh-based IBR methods.} 
Structure-from-motion (SfM)~\cite{snavely-2006-structure-from-motion} and subsequent multi-view stereo (MVS)~\cite{goesele-2007-multiviewstereo} allow for 3D reconstruction of surfaces, leading to the development of several view synthesis algorithms relying on triangle meshes as the primary 3D representation of scenes. 
Such algorithms consider textured triangles or warp and blend captured images on the mesh surface to generate novel views~\cite{wood:2000:slf,buehler2001unstructured,hedman-2018-deepblending}.
\cite{riegler2020free,riegler2021stable} consider deep learning-based mesh representations for better view synthesis, bridging the gap between traditional graphics and modern machine learning techniques.
While these mesh-based methods take advantage of existing graphics hardware and software for efficient rendering, they struggle with the capture of accurate geometry and appearance in complex regions.

\paragraph{Volumetric IBR methods.}
Volumetric methods use voxel grids, multiplane images, or neural networks to represent scenes as continuous volumetric functions of density and color.
Recently, Neural Radiance Fields~(NeRF)~\cite{mildenhall2020nerf} introduced a novel scene representation based on a continuous volumetric function parameterized by a multilayer perceptron (MLP). NeRF produces photorealistic renderings with fine details and view-dependent effects, achieved through volumetric ray tracing. However, the original NeRF is computationally expensive and memory intensive.

To address these challenges, several works have improved NeRF’s performance and scalability. These methods leverage discretized or sparse volumetric representations like voxel grids and hash tables as ways to store learnable features acting as positional encodings for 3D points~\cite{chen-eccv-2022-tensorf, karnewar2022relu, mueller2022instantngp, sun2022direct, yu2021plenoxels}, hierarchical sampling strategies~\cite{barron2022mipnerf360,hedman2021snerg,reiser2021kilonerf,yu2021plenoctrees}, or low-rank approximations~\cite{chen-eccv-2022-tensorf}. However, they still rely on volumetric ray marching, which is incompatible with standard graphics hardware and software designed for rendering polygonal surfaces.
Recent works have proposed modifying the NeRF’s representation of geometry and emitted radiance to allow for better reconstruction of specular materials~\cite{verbin2022ref} or relighting the scene through an explicit decomposition into material and lighting properties~\cite{boss2021nerd, kuang2022neroic, srinivasan2021nerv, zhang2021physg}.

\paragraph{Hybrid IBR methods.} Some methods build on differentiable rendering to combine the advantages of mesh-based and volumetric methods, and allow for surface reconstruction as well as better editability. 
They use a hybrid volume-surface representation, which enables high-quality meshes suitable for downstream graphics applications while efficiently modeling view-dependent appearance. In particular, some works optimize neural signed distance functions (SDF) by training neural radiance fields in which the density is derived as a differentiable transformation of the SDF~\cite{oechsle2021unisurf,yariv2021volsdf,wang2021neus,li-cvpr2023-neuralangelo,darmon-2022-warp,bao-2022-neumesh}. A triangle mesh can finally be reconstructed from the SDF by applying the Marching Cubes algorithm~\cite{lorensen-1987-marchingcubes}. 
However, most of these methods do not target real-time rendering. 

Alternatively, other approaches ``bake'' the rendering capacity of an optimized NeRF or neural SDF into a much efficient structure relying on an underlying triangle mesh~\cite{chen2022mobilenerf} that could benefit from the traditional triangle rasterization pipeline. 
In particular, the recent BakedSDF~\cite{yariv-2023-bakedsdf} reconstructs high quality meshes by optimizing a full neural SDF model, baking it into a high-resolution triangle mesh \vincent{that combines mesh rendering for interpolating features and deep learning to translate these features into images}, and finally optimizes a view-dependent appearance model. 

However, even though it achieves real-time rendering and produces impressive meshes of the surface of the scene, this model demands training a full neural SDF with an architecture identical to Mip-NeRF360~\cite{barron2021mipnerf}, which necessitates 48 hours of training.

Similarly, the recent method NeRFMeshing~\cite{rakotosaona2023nerfmeshing} proposes to also bake any NeRF model into a mesh structure, achieving real-time rendering. However, the meshing performed in this method lowers the quality of the rendering and results in a PSNR much lower than our method.
Additionally, this method still requires training a full NeRF model beforehand, and needs approximately an hour of training on 8 V100 NVIDIA GPUs to allow for mesh training and extraction.

\vincent{Our method is much faster at retrieveing a 3D mesh from 3D Gaussian Splatting, which is itself much faster than NeRFs. As our experiments show, our rendering done by bounding Gaussians to the mesh results in higher quality than previous solutions based on meshes.}

\paragraph{Point-based IBR methods.}  Alternatively, point-based representations for radiance field excel at modeling thin geometry and leverage fast point rasterization pipelines to render images using $\alpha$-blending rather than ray-marching~\cite{kopanas2021point, ruckert2021adop}.
In particular, the very recent 3D Gaussian Splatting model~\cite{kerbl3Dgaussians} allows for optimizing and rendering scenes with speed and quality never seen before. 

\section{3D Gaussian Splatting}
\label{sec:gaussian_splatting}

For the sake of completeness, we briefly describe the original 3D Gaussian Splatting method here. The scene is represented as a (large) set of Gaussians, where each Gaussian $g$ is represented by its mean $\mu_g$ and its covariance $\Sigma_g$ is parameterized by a scaling vector $s_g\in\IR^3$ and a quaternion $q_g\in\IR^4$ encoding the rotation of the Gaussian. In addition, each Gaussian is associated with its opacity $\alpha_g\in [0,1]$ and a set of spherical harmonics coordinates describing the colors emitted by the Gaussian for all directions. 

An image of a set of Gaussians can be rendered from a given viewpoint thanks to  a rasterizer. This rasterizer \emph{splats} the 3D Gaussians into 2D Gaussians parallel to the image plane for rendering, which results in an extremely fast rendering process.  This is the key component that makes 3D Gaussian Splatting much faster than NeRFs, as it is much faster than the ray-marching compositing required in the optimization of NeRFs.  

Given a set of images, the set of Gaussians is initialized from the point cloud produced by SfM~\cite{snavely-2006-structure-from-motion}. The Gaussians' parameters (means, quaternions, scaling vectors, but also opacities and spherical harmonics parameters) are optimized to make the renderings of the Gaussians match the input images. During optimization, more Gaussians are added to better fit the scene's geometry. As a consequence, Gaussian Splatting generally produces scenes with millions of Gaussians that can be extremely small.
\section{Method}
\label{sec:method}

\begin{figure*}[t]
  \centering
   \begin{subfigure}{0.16\linewidth}
        \includegraphics[width=\linewidth]{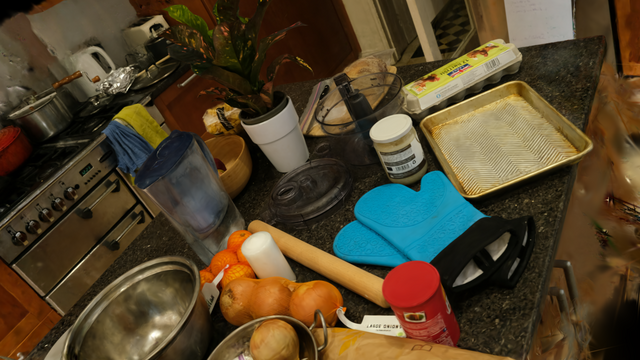}%
        \vspace{.1em}
        \includegraphics[width=\linewidth]{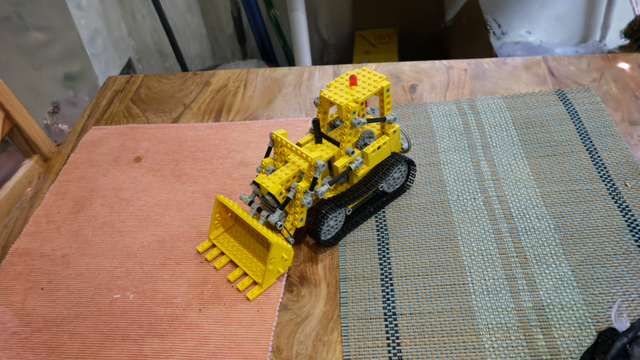}%
        \vspace{.1em}
        \includegraphics[width=\linewidth]{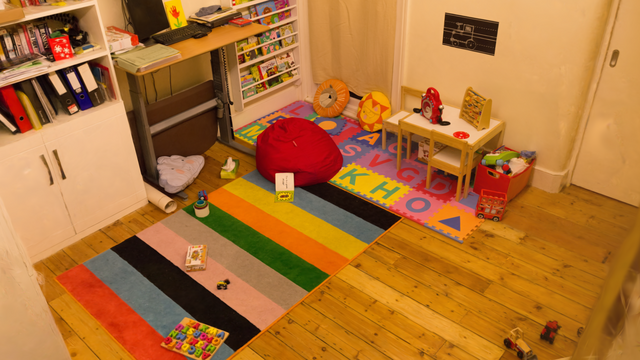}%
        \caption*{(a) Mesh \& Gaussians}
    \end{subfigure}
    \begin{subfigure}{0.16\linewidth}
        \includegraphics[width=\linewidth]{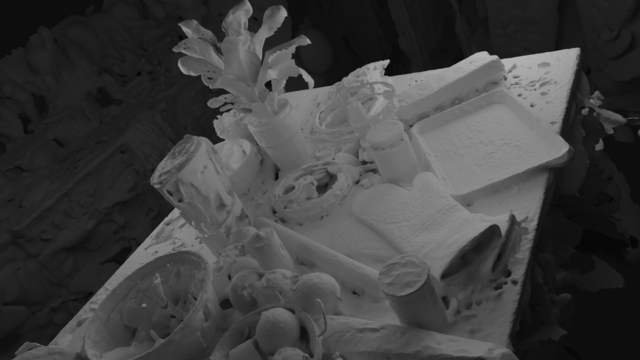}%
        \vspace{.1em}
        \includegraphics[width=\linewidth]{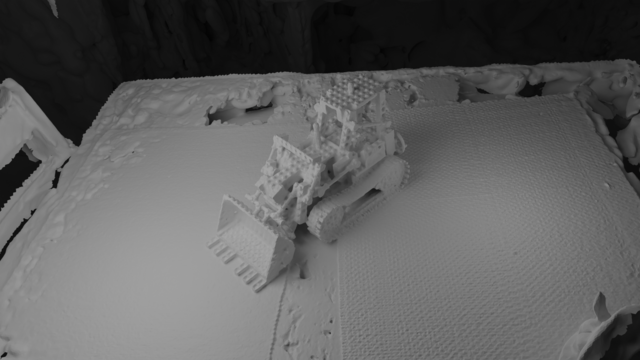}%
        \vspace{.1em}
        \includegraphics[width=\linewidth]{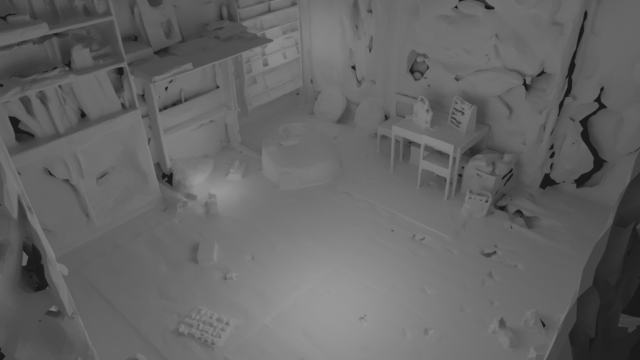}%
        \caption*{(b) Mesh (No Texture)}
    \end{subfigure}
    \begin{subfigure}{0.16\linewidth}
        \includegraphics[width=\linewidth]{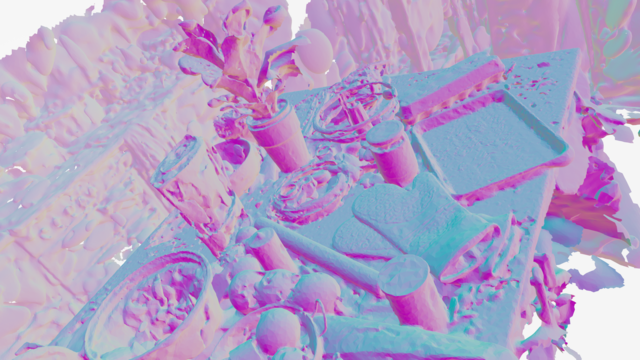}%
        \vspace{.1em}
        \includegraphics[width=\linewidth]{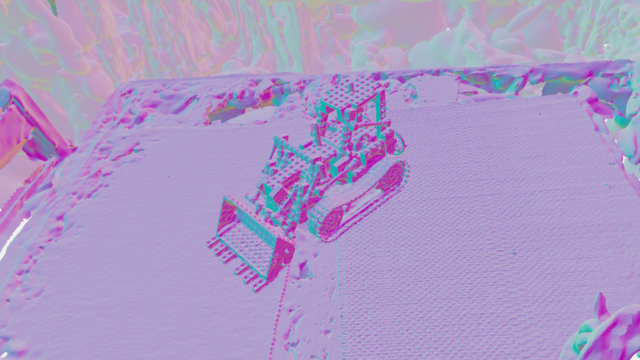}%
        \vspace{.1em}
        \includegraphics[width=\linewidth]{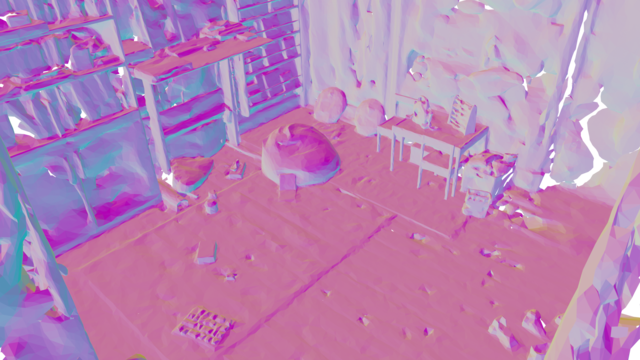}%
        \caption*{(c) Mesh normals}
    \end{subfigure}
    \hfill
    \begin{subfigure}{0.16\linewidth}
        \includegraphics[width=\linewidth]{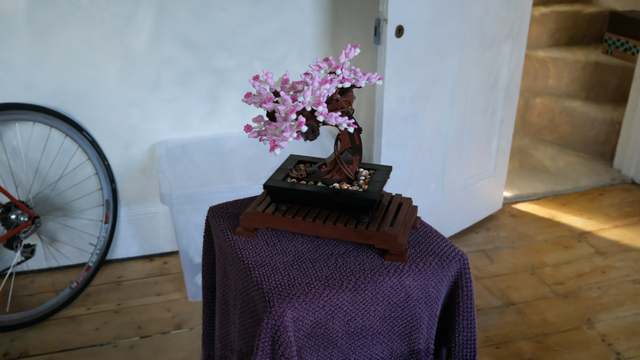}%
        \vspace{.1em}
        \includegraphics[width=\linewidth]{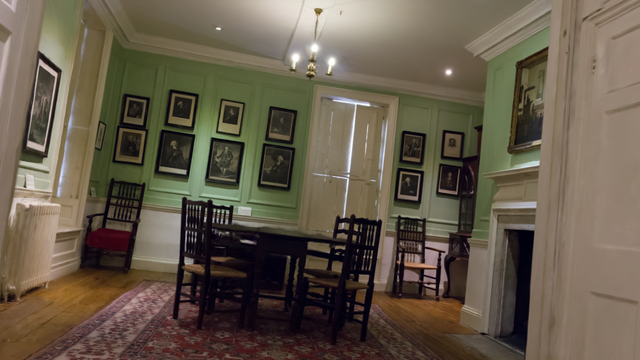}
        \vspace{.1em}
        \includegraphics[width=\linewidth]{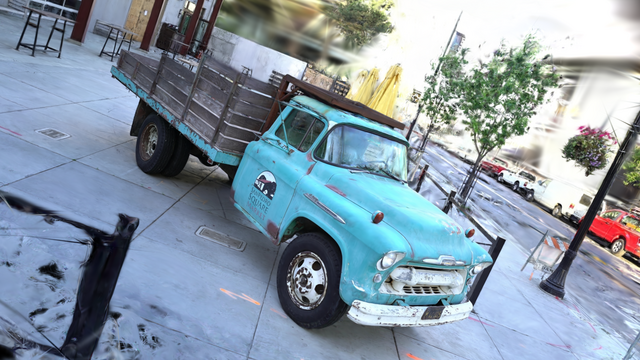}%
        \caption*{(a) Mesh \& Gaussians}
    \end{subfigure}
    \begin{subfigure}{0.16\linewidth}
        \includegraphics[width=\linewidth]{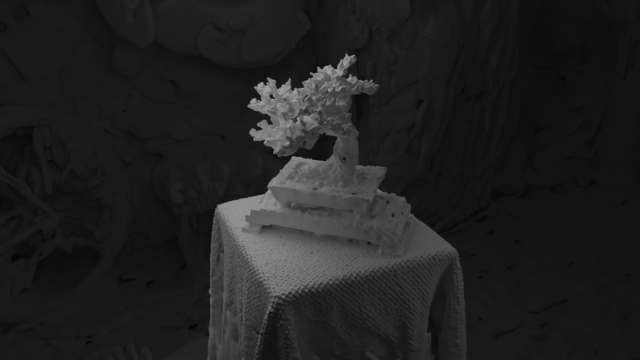}%
        \vspace{.1em}
        \includegraphics[width=\linewidth]{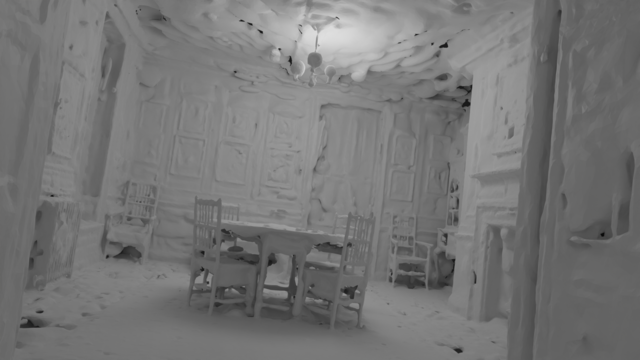}
        \vspace{.1em}
        \includegraphics[width=\linewidth]{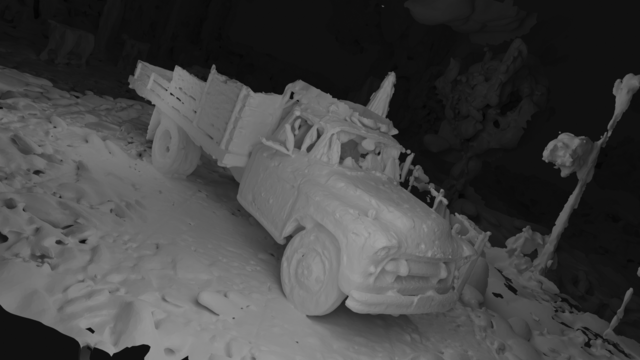}%
        \caption*{(b) Mesh (No Texture)}
    \end{subfigure}
    \begin{subfigure}{0.16\linewidth}
        \includegraphics[width=\linewidth]{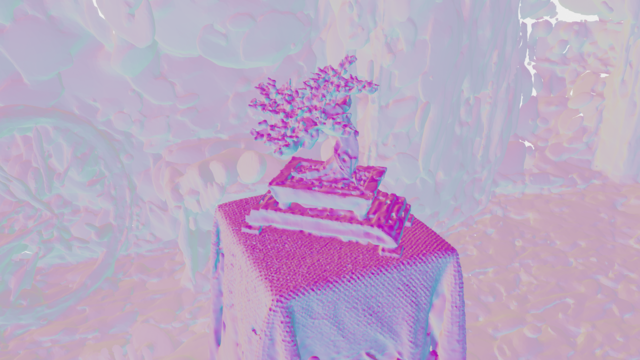}%
        \vspace{.1em}
        \includegraphics[width=\linewidth]{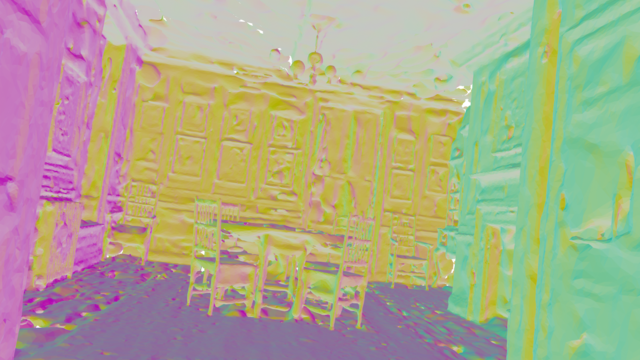}%
        \vspace{.1em}
        \includegraphics[width=\linewidth]{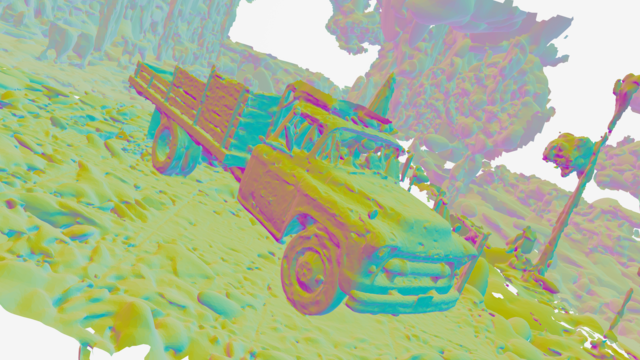}%
        \caption*{(c) Mesh normals}
    \end{subfigure}
   \caption{\textbf{Examples of (a) renderings and (b) reconstructed meshes with SuGaR.} The \textbf{(c)} normal maps help visualize the geometry.}
   \label{fig:sugar_renders}
\end{figure*}



We present our SuGaR in this section:
\begin{itemize}[noitemsep,topsep=0pt,parsep=0pt,partopsep=0pt,leftmargin=*]
\item First, we detail our loss term that enforces the alignment of the 3D Gaussians with the surface of the scene during the optimization of Gaussian Splatting. 
\item We then detail our method that exploits this alignment for extracting a highly detailed mesh from the Gaussians within minutes on a single GPU.
\item Finally, we describe our optional refinement strategy that jointly optimizes the mesh and 3D Gaussians located on the surface of the mesh using Gaussian Splatting rendering. This strategy results in a new set of Gaussians bound to an editable mesh.
\end{itemize}

\subsection{Aligning the Gaussians with the Surface}
\label{subsec:density_definition}

As discussed in the introduction, to facilitate the creation of a mesh from the Gaussians, we introduce a regularization term into the Gaussian Splatting optimization that encourages the Gaussians to be aligned with the surface of the scene and well distributed over this surface.  Our approach is to derive an SDF from the Gaussians under the assumption that the Gaussians have the desired properties. By minimizing the difference between this SDF and the  actual SDF computed for the Gaussians, we encourage the Gaussians to have these properties. 

For a given Gaussian Splatting scene, we start by considering the corresponding density function $d:\IR^3 \rightarrow \IR_+$, computed as the sum of the Gaussian values weighted by their alpha-blending coefficients at any space location $p$:
\begin{equation}
    d(p) = \sum_{g} \alpha_g \exp\left(-\frac{1}{2}(p - \mu_g)^T \Sigma^{-1}_g (p - \mu_g)\right) \> ,
    \label{eq:gaussian_splatting_density}
\end{equation}
where the $\mu_g$, $\Sigma_g$, and $\alpha_g$ are the centers, covariances, and alpha-blending coefficients of the Gaussians, respectively.
Let us consider  what  this density function becomes if the Gaussians are well distributed and aligned with the surface. 


First, in such scenario, the Gaussians would have limited overlap with their neighbors. As illustrated in Figure~\ref{fig:one}~(top-left), this is not the case in general. Then, for any point $p \in \IR^3$ close to the surface of the scene, the Gaussian $g^*$ closest to the point $p$  is likely to contribute much more than others to the density value $d(p)$. We could then approximate the Gaussian density at $p$ by:
\begin{align}
    &\alpha_{g^*} \exp\left(-\frac{1}{2}(p - \mu_{g^*})^T \Sigma^{-1}_{g^*} (p - \mu_{g^*}) \right) \> , 
    \label{eq:overlap}
\end{align}
where the ``closest Gaussian'' $g^*$ is taken as the Gaussian with the largest contribution at point $p$:
\begin{align}    
 g^* = &\arg\min_g \left\{ (p - \mu_g)^T \Sigma^{-1}_g (p - \mu_g) \right\} \> .
    \label{eq:gaussian_splatting_density_no_overlap}
\end{align}
Eq.~\eqref{eq:overlap} thus considers that the contribution of the closest Gaussian $g^*$ to the density at $p$ is much higher than the contribution of the other Gaussians. This will help us encourage the Gaussians to be well spread.

We also would like the 3D Gaussians to be flat, as they would then be aligned more closely with the surface of the mesh. Consequently, every Gaussian $g$ would have one of its three scaling factors close to 0 and:
\begin{equation}
    (p - \mu_g)^T \Sigma^{-1}_g (p - \mu_g) \approx \frac{1}{s_g^2} \langle p - \mu_g, n_g \rangle^2 \> ,
\end{equation}
where $s_g$ the smallest scaling factor of the Gaussian and $n_g$ the direction of the corresponding axis. 
%
Moreover, because we want Gaussians to describe the true surface of the scene, we need to avoid semi-transparent Gaussians. Therefore, we want Gaussians to be either opaque or fully transparent, in which case we can drop them for rendering. Consequently, we want to have $\alpha_g = 1$ for any Gaussian $g$.

In such scenario, the density of the Gaussians could finally be approximated by density $\bar{d}(p)$ with:
\begin{equation}
    \bar{d}(p) = \exp\left(-\frac{1}{2s_{g^*}^2} \langle p - \mu_{g^*}, n_{g^*} \rangle^2 \right) \> .
    \label{eq:gaussian_splatting_density_ideal}
\end{equation}

A first strategy to enforce our regularization is to add term $|d(p) - \bar{d}(p)|$ to the optimization loss. While this  approach  works well to align Gaussians with the surface,  we noticed that computing a slightly different loss relying on an SDF  rather than on density further increases the alignment of Gaussians with the surface of the scene. 
For a given flat Gaussian, i.e., $s_g = 0$, considering level sets is meaningless since all level sets would degenerate toward the plane passing through the center of the Gaussian $\mu_{g}$  with normal $n_{g}$. The distance between  point $p$ and the true surface of the scene would be approximately $|\langle p - \mu_{g'}, n_{g'} \rangle|$, the distance from $p$ to this plane. Consequently, the zero-crossings of the Signed Distance Function
\begin{equation}
    \bar{f}(p) = \pm s_{g*} \sqrt{-2 \log \left( \bar{d}(p) \right)}
    \label{eq:gaussian_splatting_sdf_ideal}
\end{equation}
corresponds to the surface of the scene. More generally, we define 
\begin{equation}
    f(p) = \pm s_{g*} \sqrt{-2 \log \left( d(p) \right)}
    \label{eq:gaussian_splatting_sdf_general}
\end{equation}
as the ``ideal'' distance function associated with the density function $d$. This distance function corresponds to the true surface of the scene in an ideal scenario where $d=\bar{d}$.
We therefore take our regularization term $\calR$ as
\begin{equation}
    \calR = \frac{1}{|\calP|} \sum_{p\in\calP} |\hat{f}(p) - f(p)| \> ,
    \label{eq:R}
\end{equation}
by sampling 3D points $p$ and summing the differences at these points between the ideal SDF $f(p)$ and an estimate $\hat{f}(p)$ of the SDF of the surface created by the current Gaussians. $\calP$  refers to the set of sampled points.

\begin{figure}
    \centering
    \includegraphics[width=0.6\linewidth]{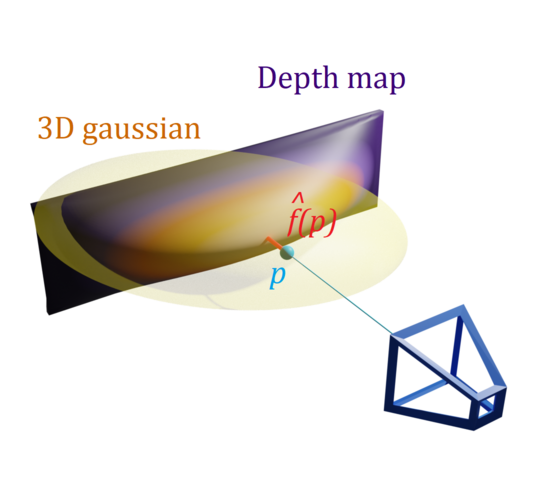}
    \vspace{-0.5cm}
    \caption{\textbf{Efficiently estimating $\hat{f}(p)$ of the SDF of the surface generated from Gaussians.} We render depth maps of the Gaussians, sample points $p$ in the viewpoint according to the distribution of the Gaussians. Value $\hat{f}(p)$ is taken as the 3D distance between $p$ and the intersection between the line of sight for $p$ and the depth map. }
    \label{fig:estimating_hat_f}
\end{figure}

Computing efficiently $\hat{f}(p)$ is \emph{a priori} challenging. To do so, we propose to use the depth maps of the Gaussians from the viewpoints used for training---these depth maps can be rendered efficiently by extending the splatting rasterizer. Then, as shown in Figure~\ref{fig:estimating_hat_f}, for a point $p$ visible from a training viewpoint, $\hat{f}(p)$ is the difference between the depth of $p$ and the depth in the corresponding depth map at the projection of $p$. Moreover, we sample points $p$ following the distribution of the Gaussians:
\begin{equation}
p \sim \prod_g \calN(. ; \mu_g, \Sigma_g) \> ,
\end{equation}
with $\calN(. ; \mu_g, \Sigma_g)$ the Gaussian distribution of mean $\mu_g$ and covariance $\Sigma_g$ as these points are likely to correspond to a high gradient for $\calR$.

We also add a regularization term to encourage the normals of SDF $f$ and the normals of SDF $\bar{f}$ to also be similar:
\begin{equation}
    \calR_\text{Norm} = \frac{1}{|\calP|} \sum_{p\in\calP}
    \left\| \frac{\nabla f(p)}{\|\nabla f(p)\|_2} - n_{g^*}\right\|_2^2 \> .
    \label{eq:R_norm}
\end{equation}

\subsection{Efficient Mesh Extraction}
\label{subsec:meshextraction}

To create a mesh from the Gaussians obtained after optimization using our regularization terms in Eq.~\eqref{eq:R} and Eq.~\eqref{eq:R_norm}, we sample 3D points on a level set of the density computed from the Gaussians.  The level set depends on a level parameter $\lambda$. Then, we obtain a mesh by simply running a Poisson reconstruction~\cite{kazhdan-2006-poissonsurfacereconstruction} on these points. Note that we can also easily assign the points with the normals of the SDF, which improves the mesh quality.

\begin{figure}
    \centering
    \begin{subfigure}{0.49\linewidth}    
    \includegraphics[height=0.9\linewidth]{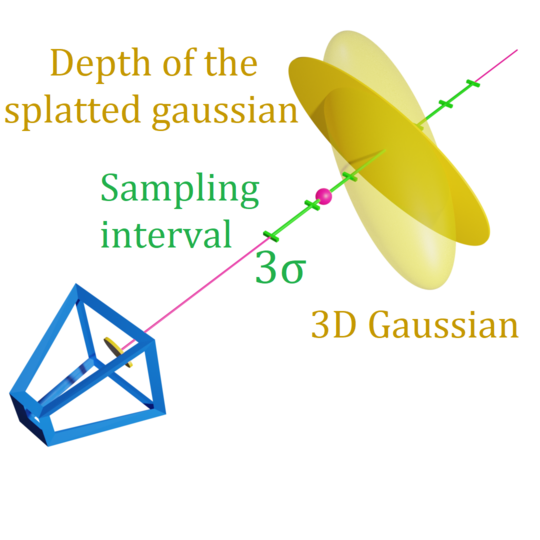}
    \end{subfigure}
    \begin{subfigure}{0.49\linewidth}    
    \hfill
    \includegraphics[trim={0.cm 0 0.cm 0},clip,width=0.49\linewidth]{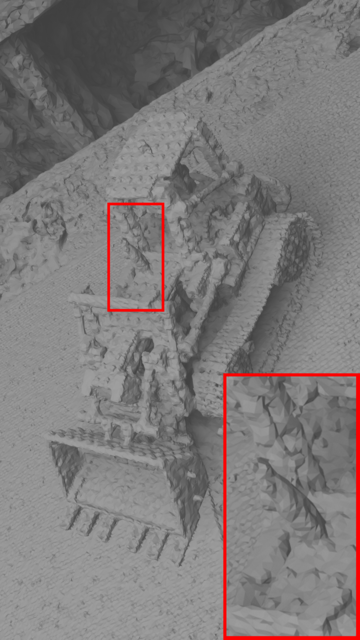}
    \hfill
    \includegraphics[trim={0.cm 0 0.cm 0},clip,width=0.49\linewidth]{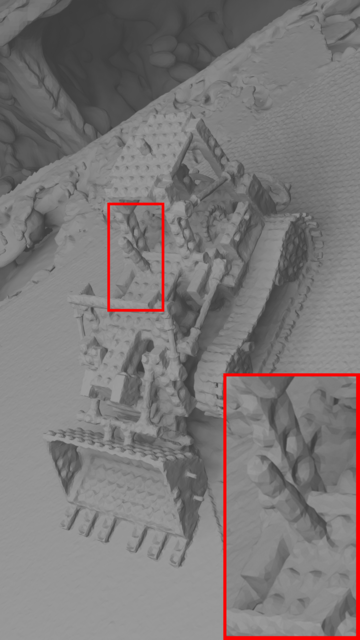}
    \end{subfigure}
    \caption{\textbf{Sampling points on a level set for Poisson reconstruction. } \textbf{Left:} We sample points on the depth maps of the Gaussians and refine the point locations to move the points on the level set. \textbf{Right:} Comparison between     
   the extracted mesh without (left) and with (right) our refinement step. Since splatted depth maps are not exact, using directly the depth points for reconstruction usually results in a large amount of noise and missing details. \vincentrmk{make the text on the left larger, make the insert as large as possible in the bottom right corner} }
    \label{fig:sampling_for_poisson}
\end{figure}

The challenge is in efficiently identifying points lying on the level set. For this, as shown in Figure~\ref{fig:sampling_for_poisson}, we again rely on the depth maps of the Gaussians as seen from the training viewpoints. We first randomly sample pixels from each depth map. For each pixel $m$, we sample its line of sight to find a 3D point on the level set. Formally, we sample $n$ points $p + t_i v$, where $p$ is the 3D point in the depth map that reprojects on pixel $m$, $v$ is the direction of the line of sight, and  $t_i \in [-3\sigma_g(v), 3\sigma_g(v)]$ where $\sigma_g(v)$ is the standard deviation of the 3D Gaussian $g$ in the direction of the camera. The interval $[-3\sigma_g(v), 3\sigma_g(v)]$ is the confidence interval for the 99.7 confidence level of the  1D Gaussian function of $t$ along the ray.

Then, we compute the density values $d_i = d(p + t_i v)$ from Eq.~\eqref{eq:gaussian_splatting_density} of these sampled points. If there exist $i,j$ such that $d_i < \lambda < d_j$, then there is a level set point located in this range. If so, we use linear interpolation to compute the coefficient $t^*$ such that $p+t^* v$ is the level set point closest to the camera, verifying $d(p+t^* v) = \lambda$. We also compute the normals of the surface at points $\hat{p}$, which we naturally define as the normalized analytical gradient of the density $\frac{\nabla d(\hat{p})}{\|\nabla d(\hat{p})\|_2}$.

Finally, we apply Poisson reconstruction to reconstruct a surface mesh from the level set points and their normals. 

\subsection{Binding New 3D Gaussians to the Mesh}
\label{subsec:jointrefinement}

Once we have extracted a first mesh, we can refine this mesh by binding new Gaussians to the mesh triangles and optimize the Gaussians and the mesh jointly using the Gaussian Splatting rasterizer. This enables the edition of the Gaussian splatting scene with popular mesh editing tools while keeping high-quality rendering thanks to the Gaussians. 

\begin{figure}
    \centering
     \begin{subfigure}{0.49\linewidth}    
    \hfill
    \includegraphics[trim={0.cm 0 0.cm 0},clip,width=0.9\linewidth]{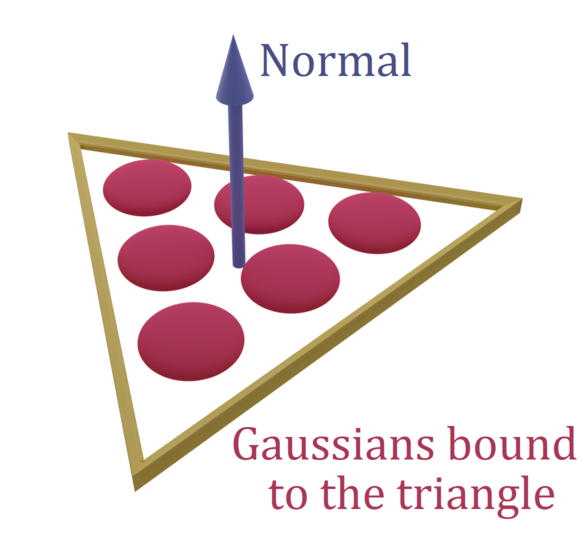}
    \end{subfigure}
    \hfill
    \begin{subfigure}{0.49\linewidth}    
    \includegraphics[trim={0.cm 0 0.cm 0},clip,width=0.9\linewidth]{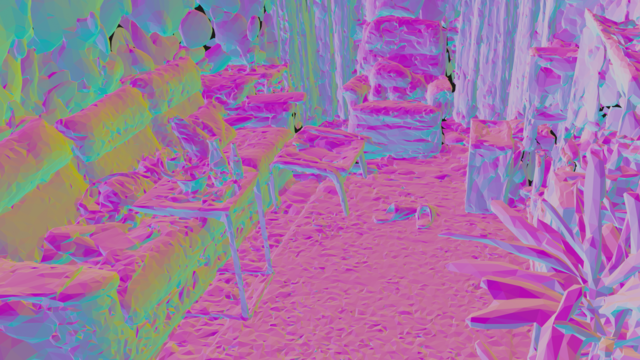}
    \hfill
    \includegraphics[trim={0.cm 0 0.cm 0},clip,width=0.9\linewidth]{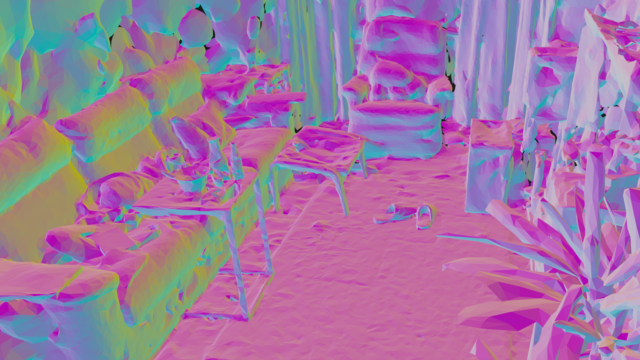}
    \end{subfigure}
    \caption{\textbf{Joint refinement of mesh and Gaussians. }
    \textbf{Left:} We bind Gaussians to the triangles of the mesh. Depending on the number of triangles in the scene, we bind a different number of Gaussians per triangle, with predefined barycentric coordinates. \textbf{Right:} Mesh before and after joint refinement.}
    \label{fig:jointrefinement}
\end{figure}

Given the initial mesh, we instantiate new 3D Gaussians on the mesh. More exactly,  we associate a set of $n$ thin 3D Gaussians to each triangle of the mesh, sampled on the surface of the triangle, as illustrated in Figure~\ref{fig:jointrefinement}. To do so, we slightly modify the structure of the original 3D Gaussian Splatting model.

We explicitly compute the means of the Gaussians from the mesh vertices using predefined barycentric coordinates in the corresponding triangles during optimization. Also, the Gaussians have only 2 learnable scaling factors instead of 3 and only 1 learnable 2D rotation encoded with a complex number rather than a quaternion, to keep the Gaussians flat and aligned with the mesh triangles. More details about this parameterisation are given in the supplementary material. Like the original model, we also optimize an opacity value and a set of spherical harmonics for every Gaussian to encode the color emitted in all directions.

Figure~\ref{fig:jointrefinement} shows an example of a mesh before and after refinement. Figure~\ref{fig:teaser_editing} and the supplementary material give examples of what can be done by editing the mesh.
\section{Experiments}
\label{sec:experiments}

\subsection{Implementation details}

All our models are optimized on a single GPU Nvidia Tesla V100 SXM2 32 Go. 

\paragraph{Regularization.} For all scenes, we start by optimizing a Gaussian Splatting with no regularization for 7,000 iterations in order to let the 3D Gaussians position themselves without any additional constraint.
Then, we perform 2,000 iterations with an additional entropy loss on the opacities $\alpha_g$ of the Gaussians, as a way to enforce them to become binary.

Finally, we remove Gaussians with opacity values under 0.5 and perform 6,000 iterations with the regularization term introduced in Subsection~\ref{subsec:density_definition}, which makes a total of 15,000 iterations. To compute the density values of points from a Gaussian $g$, we sum only the Gaussian functions from the 16 nearest Gaussians of $g$ and update the list of nearest neighbors every 500 iterations. Optimization typically takes between 15 and 45 minutes depending on the scene.

\paragraph{Mesh extraction.} For all experiments except the ablation presented in Table~\ref{tab:mesh_extraction}, we extract the $\lambda$-level set of the density function for $\lambda = 0.3$. We perform Poisson reconstruction with depth 10 and apply mesh simplification using quadric error metrics~\cite{garland-1997-quadricerrormetricsdecimation} to decrease the resolution of the meshes. Mesh extraction generally takes between 5 and 10 minutes depending on the scene.

\paragraph{Joint refinement.} We jointly refine the mesh and the bound 3D Gaussians for either 2,000, 7,000 or 15,000 iterations. Depending on the number of iterations, the duration of refinement goes from a few minutes to an hour.

\subsection{Real-Time Rendering of Real Scenes}

\begin{table*}
  \centering
  {\small
  \begin{tabular}{@{}lcccccccccc@{}}
    \toprule
      %
     %
     \multicolumn{1}{c}{} & \multicolumn{3}{c}{Indoor scenes} & \multicolumn{3}{c}{Outdoor scenes} & \multicolumn{3}{c}{Average on all scenes} \\
     \cmidrule(r){2-4} \cmidrule(r){5-7} \cmidrule(r){8-10}
      & PSNR $\uparrow$ & SSIM $\uparrow$ & LPIPS $\downarrow$ & PSNR $\uparrow$ & SSIM $\uparrow$ & LPIPS $\downarrow$ & PSNR $\uparrow$ & SSIM $\uparrow$ & LPIPS $\downarrow$ \\
    \midrule
    \multicolumn{10}{l}{\textbf{No mesh (except SuGaR)}} \\
    \midrule
    Plenoxels~\cite{yu_and_fridovichkeil2021plenoxels} & 24.83 & 0.766 & 0.426 & 22.02 & 0.542 & 0.465 & 23.62 & 0.670 & 0.443 \\
    INGP-Base~\cite{mueller2022instantngp} & 28.65 & 0.840 & 0.281 & 23.47 & 0.571 & 0.416 & 26.43 & 0.725 & 0.339 \\
    INGP-Big~\cite{mueller2022instantngp} & 29.14 & 0.863 & 0.242 & 23.57 & 0.602 & 0.375 & 26.75 & 0.751 & 0.299 \\
    Mip-NeRF360~\cite{barron2022mipnerf360} & \cellcolor{red!25}\textbf{31.58} & \cellcolor{orange!25}0.914 & \cellcolor{red!25}\textbf{0.182} & \cellcolor{orange!25}25.79 & \cellcolor{orange!25}0.746 & \cellcolor{orange!25}0.247 & \cellcolor{red!25}\textbf{29.09} & \cellcolor{orange!25}0.842 & \cellcolor{orange!25}0.210 \\
    3DGS~\cite{kerbl3Dgaussians} & \cellcolor{orange!25}30.41 & \cellcolor{red!25}\textbf{0.920} & \cellcolor{orange!25}0.189 & \cellcolor{red!25}\textbf{26.40} & \cellcolor{red!25}\textbf{0.805} & \cellcolor{red!25}\textbf{0.173} & \cellcolor{orange!25}28.69 & \cellcolor{red!25}\textbf{0.870} & \cellcolor{red!25}\textbf{0.182} \\
    R-SuGaR-15K (Ours) & \cellcolor{yellow!25}29.43 & \cellcolor{yellow!25}0.910 & \cellcolor{yellow!25}0.216 & \cellcolor{yellow!25}24.40 & \cellcolor{yellow!25}0.699 & \cellcolor{yellow!25}0.301 & \cellcolor{yellow!25}27.27 & \cellcolor{yellow!25}0.820 & \cellcolor{yellow!25}0.253 \\
    \midrule
    \multicolumn{10}{l}{\textbf{With mesh}} \\
    \midrule
    Mobile-NeRF~\cite{chen2022mobilenerf} & -- & -- & -- & 21.95 & 0.470 & 0.470 & -- & -- & -- \\
    NeRFMeshing~\cite{rakotosaona2023nerfmeshing} & 23.83 & -- & -- & 22.23 & -- & -- & 23.15 & -- & -- \\
    BakedSDF~\cite{yariv-2023-bakedsdf} & \cellcolor{yellow!25}27.06 & 0.836 & \cellcolor{yellow!25}0.258 & -- & -- & -- & -- & -- & -- \\
    R-SuGaR-2K (Ours) & 26.29 & \cellcolor{yellow!25}0.872 & 0.262 & \cellcolor{yellow!25}22.97 & \cellcolor{yellow!25}0.648 & \cellcolor{yellow!25}0.360 & \cellcolor{yellow!25}24.87 & \cellcolor{yellow!25}0.776 & \cellcolor{yellow!25}0.304 \\
    R-SuGaR-7K (Ours) & \cellcolor{orange!25}28.73 & \cellcolor{orange!25}0.904 & \cellcolor{orange!25}0.226 & \cellcolor{orange!25}24.16 & \cellcolor{orange!25}0.691 & \cellcolor{orange!25}0.313 & \cellcolor{orange!25}26.77 & \cellcolor{orange!25}0.813 & \cellcolor{orange!25}0.263 \\
    R-SuGaR-15K (Ours) & \cellcolor{red!25}\textbf{29.43} & \cellcolor{red!25}\textbf{0.910} & \cellcolor{red!25}\textbf{0.216} & \cellcolor{red!25}\textbf{24.40} & \cellcolor{red!25}\textbf{0.699} & \cellcolor{red!25}\textbf{0.301} & \cellcolor{red!25}\textbf{27.27} & \cellcolor{red!25}\textbf{0.820} & \cellcolor{red!25}\textbf{0.253} \\
    \bottomrule
  \end{tabular}
  }
  \caption{\textbf{Quantitative evaluation of rendering quality on the Mip-NeRF360 dataset~\cite{barron2022mipnerf360}.} SuGaR is best among the methods that recover a mesh, and still performs well compared to NeRF methods and vanilla 3D Gaussian Splatting. }
  \label{tab:nvsmetrics_mipnerf360}
\end{table*}




For evaluating our model, we follow the approach from the original 3D Gaussian Splatting paper~\cite{kerbl3Dgaussians} and compare the performance of several variations of our method SuGaR after refinement on real 3D scenes from 3 different datasets: Mip-NeRF360~\cite{barron2021mipnerf}, DeepBlending~\cite{hedman-2018-deepblending} and Tanks\&Temples~\cite{knapitsch-2017-tanksandtemples}. We call R-SuGaR-\textit{N}K a refined SuGaR model optimized for \textit{N} iterations during refinement.

Following~\cite{kerbl3Dgaussians}, we select the same sets of 2 scenes from Tanks\&Temples~(\textit{Truck} and \textit{Train}) and 2 scenes from DeepBlending~(\textit{Playroom} and \textit{Dr. Johnson}). However, due to licensing issues and the unavailability of the scenes \textit{Flowers} and \textit{Treehill}, we perform the evaluation of all methods only on 7 scenes from Mip-NeRF360 instead of the full set of 9 scenes.

We compute the standard metrics PSNR, SSIM and LPIPS~\cite{zhang2018lpips} to evaluate the quality of SuGaR's rendering using our extracted meshes and their bound surface Gaussians. \vincent{Note that \cite{chen2022mobilenerf, yariv-2023-bakedsdf, rakotosaona2023nerfmeshing} also do not use plain textured mesh rendering.} \vincentrmk{ but .. what do they do?} We compare to several baselines, some of them focusing only on Novel View Synthesis~\cite{yu2021plenoxels, mueller2022instantngp, barron2022mipnerf360, kerbl3Dgaussians} and others relying on a reconstructed mesh~\cite{chen2022mobilenerf, yariv-2023-bakedsdf, rakotosaona2023nerfmeshing}, just like our method SuGaR. 
Results on the Mip-NeRF360 dataset are given in Table~\ref{tab:nvsmetrics_mipnerf360}. Results on Tanks\&Temple and DeepBlending are similar and can be found in the supplementary material.


Even though SuGaR focuses on aligning 3D Gaussians for reconstructing a high quality mesh during the first stage of its optimization, it significantly outperforms the state of the art methods for Novel View Synthesis using a mesh and reaches better performance than several famous models that focus only on rendering, such as Instant-NGP~\cite{mueller2022instantngp} and Plenoxels~\cite{yu2021plenoxels}. This performance is remarkable as SuGaR is able to extract a mesh significantly faster than other methods.

Moreover, SuGaR even reaches performance similar to state-of-the-art models for rendering quality~\cite{barron2022mipnerf360, kerbl3Dgaussians} on some of the scenes used for evaluation. Two main reasons explain this performance. First, the mesh extracted after the first stage of optimization serves as an excellent initialization for positioning Gaussians when starting the refinement phase. Then, the Gaussians constrained to remain on the surface during refinement greatly increase the rendering quality as they play the role of an efficient texturing tool and help reconstructing very fine details missing in the extracted mesh. 
Additional qualitative results are available in Figure~\ref{fig:sugar_renders}.

\begin{table}
  \centering
  {\small
  \begin{tabular}{@{}lcccc@{}}
    \toprule
      %
     %
     Extraction method & PSNR $\uparrow$ & SSIM $\uparrow$ & LPIPS $\downarrow$ \\
    \midrule
     Marching Cubes~\cite{lorensen-1987-marchingcubes} & 23.91 & 0.703 & 0.392 \\
     Poisson (centers)~\cite{kazhdan-2006-poissonsurfacereconstruction} & 23.76 & 0.756 & \cellcolor{yellow!25}0.340 \\
     Ours (Surface level 0.1) & \cellcolor{yellow!25}24.62 & \cellcolor{yellow!25}0.765 & \cellcolor{orange!25}0.313 \\ 
     Ours (Surface level 0.3) & \cellcolor{orange!25}24.87 & \cellcolor{orange!25}\textbf{0.776} & \cellcolor{red!25}\textbf{0.304} \\
     Ours (Surface level 0.5) & \cellcolor{red!25}\textbf{24.91} & \cellcolor{red!25}\textbf{0.777} & \cellcolor{red!25}\textbf{0.304} \\
    \bottomrule
  \end{tabular}
  }
  \caption{\textbf{Ablation for different mesh extraction methods on the Mip-NeRF360 dataset~\cite{barron2022mipnerf360} \vincent{after applying our regularization term}.}   For 'Poisson (centers)', we apply Poisson reconstruction~\cite{kazhdan-2006-poissonsurfacereconstruction} using as surface points the centers of the 3D Gaussians.  For fair comparison, we calibrate the methods to enforce all extracted meshes to have approximately 1,000,000 vertices.
  }
  \label{tab:mesh_extraction}
\end{table}
\begin{table}
  \centering
  {\small
  \begin{tabular}{@{}lcccc@{}}
    \toprule
      %
     %
      & PSNR $\uparrow$ & SSIM $\uparrow$ & LPIPS $\downarrow$ \\
    \midrule
     1M vertices (3DGS) & \cellcolor{red!25}\textbf{24.51} & \cellcolor{red!25}\textbf{0.768} & \cellcolor{red!25}\textbf{0.295} \\
     1M vertices (UV) & 21.24 & 0.609 & 0.478 \\
     \midrule
     200K vertices (3DGS) & \cellcolor{orange!25}24.24 & 0.\cellcolor{orange!25}757 & \cellcolor{orange!25}0.300 \\
     200K vertices (UV) & \cellcolor{yellow!25}21.44 & \cellcolor{yellow!25}0.656 & \cellcolor{yellow!25}0.419 \\
    \bottomrule
  \end{tabular}
  }
  \caption{\textbf{Comparison between surface-aligned 3D Gaussians and an optimized traditional UV texture on the Mip-NeRF360 dataset~\cite{barron2022mipnerf360}}. 
  For fair comparison, we only use the diffuse spherical harmonics component when rendering images with SuGaR. Using 3D Gaussians bound to the mesh greatly improves rendering quality, even though it contains less parameters than the UV texture.
  }
  \label{tab:traditionaltexture}
\end{table}


\subsection{Mesh Extraction}
To demonstrate the ability of our mesh extraction method for reconstructing high-quality meshes that are well-suited for view synthesis, we compare different mesh extraction algorithms. 
In particular, we optimize several variations of SuGaR by following the exact same pipeline as our standard model, except for the mesh extraction process: We either extract the mesh using a very fine marching cubes algorithm~\cite{lorensen-1987-marchingcubes}, by applying Poisson reconstruction~\cite{kazhdan-2006-poissonsurfacereconstruction} using the centers of the 3D Gaussians as the surface point cloud, or by applying our mesh extraction method on different level sets.
Quantitative results are available in Table~\ref{tab:mesh_extraction} and show the clear superiority of our approach for meshing 3D Gaussians. Figure~\ref{fig:one} also illustrates how the marching cubes algorithm fails in this context.



\subsection{Mesh Rendering Ablation} 

Table~\ref{tab:traditionaltexture} provides additional results to quantify how various parameters impact rendering performance.
In particular, we evaluate how the resolution of the mesh extraction, i.e., the number of triangles, modifies the rendering quality. For fair comparison, we increase the number of surface-aligned Gaussians per triangle when we decrease the number of triangles. Results show that increasing the number of vertices increases the quality of rendering with surface Gaussians, but meshes with less triangles are already able to reach state of the art results.

Then, we illustrate the benefits of using Gaussians aligned on the surface as a texturing tool for rendering meshes. To this end, we also optimize traditional UV textures on our meshes using differentiable mesh rendering with traditional triangle rasterization. Even though rendering with surface-aligned Gaussians provides better performance, rendering our meshes with traditional UV textures still produces satisfying results, which further illustrates the quality of our extracted meshes. 
Qualitative comparisons are provided in the supplementary material.


\section{Conclusion}
\label{sec:conclusion}

We proposed a very fast algorithm to obtain an accurate 3D triangle mesh for a scene via Gaussian Splatting. Moreover, by combining meshing and Gaussian Splatting, we make possible intuitive manipulation of the captured scenes and realistic rendering, offering new possibilities for creators.
}

\section*{Acknowledgements.}
\label{sec:acknowledgements}

This work was granted access to the HPC resources of IDRIS under the allocation 2023-AD011013387R1 made by GENCI. We thank George Drettakis and Elliot Vincent for inspiring discussions and valuable feedback.
{
    \small
    \bibliographystyle{ieeenat_fullname}
    \bibliography{arxiv}

\begin{thebibliography}{44}
\providecommand{\natexlab}[1]{#1}
\providecommand{\url}[1]{\texttt{#1}}
\expandafter\ifx\csname urlstyle\endcsname\relax
  \providecommand{\doi}[1]{doi: #1}\else
  \providecommand{\doi}{doi: \begingroup \urlstyle{rm}\Url}\fi

\bibitem[Barron(2021)]{barron2021mipnerf}
Jonathan~T. Barron.
\newblock {{Mip-NeRF}: A Multiscale Representation for Anti-Aliasing Neural Radiance Fields}.
\newblock In \emph{International Conference on Computer Vision}, 2021.

\bibitem[Barron(2022)]{barron2022mipnerf360}
Jonathan~T. Barron.
\newblock {Mip-NeRF 360: Unbounded Anti-Aliased Neural Radiance Fields}.
\newblock In \emph{Conference on Computer Vision and Pattern Recognition}, 2022.

\bibitem[Boss et~al.(2021)Boss, Braun, Jampani, Barron, Liu, and Lensch]{boss2021nerd}
Mark Boss, Raphael Braun, Varun Jampani, Jonathan~T. Barron, Ce Liu, and Hendrik P.~A. Lensch.
\newblock {{NeRD}: Neural Reflectance Decomposition from Image Collections}.
\newblock In \emph{International Conference on Computer Vision}, 2021.

\bibitem[Buehler et~al.(2001)Buehler, Bosse, Mcmillan, Gortler, and Cohen]{buehler2001unstructured}
Chris Buehler, Michael Bosse, Leonard Mcmillan, Steven Gortler, and Michael Cohen.
\newblock {Unstructured Lumigraph Rendering}.
\newblock In \emph{ACM SIGGRAPH}, 2001.

\bibitem[Chen et~al.(2022)Chen, Xu, Geiger, Yu, and Su]{chen-eccv-2022-tensorf}
Anpei Chen, Zexiang Xu, Andreas Geiger, Jingyi Yu, and Hao Su.
\newblock {{TensoRF}: Tensorial Radiance Fields}.
\newblock In \emph{European Conference on Computer Vision}, 2022.

\bibitem[Chen et~al.(2023)Chen, Funkhouser, Hedman, and Tagliasacchi]{chen2022mobilenerf}
Zhiqin Chen, Thomas Funkhouser, Peter Hedman, and Andrea Tagliasacchi.
\newblock {MobileNeRF: Exploiting the Polygon Rasterization Pipeline for Efficient Neural Field Rendering on Mobile Architectures}.
\newblock In \emph{Conference on Computer Vision and Pattern Recognition}, 2023.

\bibitem[{Chong Bao and Bangbang Yang} et~al.(2022){Chong Bao and Bangbang Yang}, Junyi, Hujun, Yinda, Zhaopeng, and Guofeng]{bao-2022-neumesh}
{Chong Bao and Bangbang Yang}, Zeng Junyi, Bao Hujun, Zhang Yinda, Cui Zhaopeng, and Zhang Guofeng.
\newblock {NeuMesh: Learning Disentangled Neural Mesh-Based Implicit Field for Geometry and Texture Editing}.
\newblock In \emph{European Conference on Computer Vision}, 2022.

\bibitem[Darmon et~al.(2022)Darmon, Bascle, Devaux, Monasse, and Aubry]{darmon-2022-warp}
Fran{\c{c}}ois Darmon, B\'en\'edicte Bascle, Jean-Cl\'ement Devaux, Pascal Monasse, and Mathieu Aubry.
\newblock {Improving Neural Implicit Surfaces Geometry with Patch Warping}.
\newblock In \emph{Conference on Computer Vision and Pattern Recognition}, 2022.

\bibitem[Garland and Heckbert(1997)]{garland-1997-quadricerrormetricsdecimation}
Michael Garland and Paul~S. Heckbert.
\newblock {Surface Simplification Using Quadric Error Metrics}.
\newblock In \emph{ACM SIGGRAPH}, 1997.

\bibitem[Goesele et~al.(2007)Goesele, Snavely, Curless, Hoppe, and Seitz]{goesele-2007-multiviewstereo}
Michael Goesele, Noah Snavely, Brian Curless, Hugues Hoppe, and Steven Seitz.
\newblock {Multi-View Stereo for Community Photo Collections}.
\newblock In \emph{International Conference on Computer Vision}, 2007.

\bibitem[Hedman and Srinivasan(2021)]{hedman2021snerg}
Peter Hedman and Pratul~P. Srinivasan.
\newblock {Baking Neural Radiance Fields for Real-Time View Synthesis}.
\newblock In \emph{International Conference on Computer Vision}, 2021.

\bibitem[Hedman et~al.(2018)Hedman, Philip, Price, Frahm, Drettakis, and Brostow]{hedman-2018-deepblending}
Peter Hedman, Julien Philip, True Price, Jan-Michael Frahm, George Drettakis, and Gabriel Brostow.
\newblock {Deep Blending for Free-Viewpoint Image-Based Rendering}.
\newblock In \emph{ACM SIGGRAPH}, 2018.

\bibitem[Karnewar et~al.(2022)Karnewar, Ritschel, Wang, and Mitra]{karnewar2022relu}
Animesh Karnewar, Tobias Ritschel, Oliver Wang, and Niloy Mitra.
\newblock {{ReLU} Fields: The Little Non-Linearity That Could}.
\newblock In \emph{ACM SIGGRAPH}, 2022.

\bibitem[Kazhdan et~al.(2006)Kazhdan, Bolitho, and Hoppe]{kazhdan-2006-poissonsurfacereconstruction}
Michael~M. Kazhdan, Matthew Bolitho, and Hugues Hoppe.
\newblock {Poisson Surface Reconstruction}.
\newblock In \emph{Eurographics}, 2006.

\bibitem[Kerbl et~al.(2023)Kerbl, Kopanas, Leimk\"uhler, and Drettakis]{kerbl3Dgaussians}
Bernhard Kerbl, Georgios Kopanas, Thomas Leimk\"uhler, and George Drettakis.
\newblock {3D Gaussian Splatting for Real-Time Radiance Field Rendering}.
\newblock In \emph{ACM SIGGRAPH}, 2023.

\bibitem[Knapitsch et~al.(2017)Knapitsch, Park, Zhou, and Koltun]{knapitsch-2017-tanksandtemples}
Arno Knapitsch, Jaesik Park, Qian-Yi Zhou, and Vladlen Koltun.
\newblock {Tanks and Temples: Benchmarking Large-Scale Scene Reconstruction}.
\newblock In \emph{ACM SIGGRAPH}, 2017.

\bibitem[Kopanas et~al.(2021)Kopanas, Philip, Leimk\"uhler, and Drettakis]{kopanas2021point}
Georgios Kopanas, Julien Philip, Thomas Leimk\"uhler, and George Drettakis.
\newblock {Point-Based Neural Rendering with Per-View Optimization}.
\newblock In \emph{Computer Graphics Forum}, 2021.

\bibitem[Kuang et~al.(2022)Kuang, Olszewski, Chai, Huang, Achlioptas, and Tulyakov]{kuang2022neroic}
Zhengfei Kuang, Kyle Olszewski, Menglei Chai, Zeng Huang, Panos Achlioptas, and Sergey Tulyakov.
\newblock {{NeROIC}: Neural Rendering of Objects from Online Image Collections}.
\newblock In \emph{ACM SIGGRAPH}, 2022.

\bibitem[Levoy and Hanrahan(1996)]{levoy1996light}
Marc Levoy and Pat Hanrahan.
\newblock {Light Field Rendering}.
\newblock In \emph{ACM SIGGRAPH}, 1996.

\bibitem[Li et~al.(2023)Li, M\"uller, Evans, Taylor, Unberath, Liu, and Lin]{li-cvpr2023-neuralangelo}
Zhaoshuo Li, Thomas M\"uller, Alex Evans, Russell~H. Taylor, Mathias Unberath, Ming-Yu Liu, and Chen-Hsuan Lin.
\newblock {Neuralangelo: High-Fidelity Neural Surface Reconstruction}.
\newblock In \emph{Conference on Computer Vision and Pattern Recognition}, 2023.

\bibitem[Lorensen and Cline(1987)]{lorensen-1987-marchingcubes}
William~E. Lorensen and Harvey~E. Cline.
\newblock {Marching Cubes: A High Resolution {3D} Surface Construction Algorithm}.
\newblock In \emph{ACM SIGGRAPH}, 1987.

\bibitem[Mildenhall et~al.(2020)Mildenhall, Srinivasan, Tancik, Barron, Ramamoorthi, and Ng]{mildenhall2020nerf}
Ben Mildenhall, Pratul~P. Srinivasan, Matthew Tancik, Jonathan~T. Barron, Ravi Ramamoorthi, and Ren Ng.
\newblock {NeRF: Representing Scenes as Neural Radiance Fields for View Synthesis}.
\newblock In \emph{European Conference on Computer Vision}, 2020.

\bibitem[M\"uller et~al.(2022)M\"uller, Evans, Schied, and Keller]{mueller2022instantngp}
Thomas M\"uller, Alex Evans, Christoph Schied, and Alexander Keller.
\newblock {Instant Neural Graphics Primitives with a Multiresolution Hash Encoding}.
\newblock In \emph{ACM SIGGRAPH}, 2022.

\bibitem[Oechsle et~al.(2021)Oechsle, Peng, and Geiger]{oechsle2021unisurf}
Michael Oechsle, Songyou Peng, and Andreas Geiger.
\newblock {{UNISURF}: Unifying Neural Implicit Surfaces and Radiance Fields for Multi-View Reconstruction}.
\newblock In \emph{International Conference on Computer Vision}, 2021.

\bibitem[Paszke et~al.(2019)Paszke, Gross, Massa, Lerer, Bradbury, Chanan, Killeen, Lin, Gimelshein, Antiga, Desmaison, Kopf, Yang, Devito, Raison, Tejani, Chilamkurthy, Steiner, Fang, Bai, and Chintala]{paszke-nips19-pytorch}
Adam Paszke, Sam Gross, Francisco Massa, Adam Lerer, James Bradbury, Gregory Chanan, Trevor Killeen, Zeming Lin, Natalia Gimelshein, Luca Antiga, Alban Desmaison, Andreas Kopf, Edward Yang, Zachary Devito, Martin Raison, Alykhan Tejani, Sasank Chilamkurthy, Benoit Steiner, Lu Fang, Junjie Bai, and Soumith Chintala.
\newblock {PyTorch: An Imperative Style, High-Performance Deep Learning Library}.
\newblock In \emph{Advances in Neural Information Processing Systems}. Curran Associates Inc., 2019.

\bibitem[Rakotosaona et~al.(2023)Rakotosaona, Manhardt, Arroyo, Niemeyer, Kundu, and Tombari]{rakotosaona2023nerfmeshing}
Marie-Julie Rakotosaona, Fabian Manhardt, Diego~Martin Arroyo, Michael Niemeyer, Abhijit Kundu, and Federico Tombari.
\newblock {NeRFMeshing: Distilling Neural Radiance Fields into Geometrically-Accurate 3D Meshes}.
\newblock In \emph{DV}, 2023.

\bibitem[Ravi et~al.(2020)Ravi, Reizenstein, Novotny, Gordon, Lo, Johnson, and Gkioxari]{ravi-arxiv20-accelerating3ddeeplearning}
Nikhila Ravi, Jeremy Reizenstein, David Novotny, Taylor Gordon, Wan-Yen Lo, Justin Johnson, and Georgia Gkioxari.
\newblock {Accelerating 3D Deep Learning with PyTorch3D}.
\newblock In \emph{arXiv Preprint}, 2020.

\bibitem[Reiser et~al.(2021)Reiser, Peng, Liao, and Geiger]{reiser2021kilonerf}
Christian Reiser, Songyou Peng, Yiyi Liao, and Andreas Geiger.
\newblock {{KiloNeRF}: Speeding Up Neural Radiance Fields with Thousands of Tiny {MLPs}}.
\newblock In \emph{International Conference on Computer Vision}, 2021.

\bibitem[Riegler and Koltun(2020)]{riegler2020free}
Gernot Riegler and Vladlen Koltun.
\newblock {Free View Synthesis}.
\newblock In \emph{European Conference on Computer Vision}, 2020.

\bibitem[Riegler and Koltun(2021)]{riegler2021stable}
Gernot Riegler and Vladlen Koltun.
\newblock {Stable View Synthesis}.
\newblock In \emph{Conference on Computer Vision and Pattern Recognition}, 2021.

\bibitem[R\"uckert et~al.(2022)R\"uckert, Franke, and Stamminger]{ruckert2021adop}
Darius R\"uckert, Linus Franke, and Marc Stamminger.
\newblock {{ADOP}: Approximate Differentiable One-Pixel Point Rendering}.
\newblock In \emph{ACM SIGGRAPH}, 2022.

\bibitem[Snavely et~al.(2006)Snavely, Seitz, and Szeliski]{snavely-2006-structure-from-motion}
Noah Snavely, Steven~M. Seitz, and Richard Szeliski.
\newblock {Photo Tourism: Exploring Photo Collections in 3D}.
\newblock In \emph{ACM SIGGRAPH}, 2006.

\bibitem[Srinivasan et~al.(2021)Srinivasan, Deng, Zhang, Tancik, Mildenhall, and Barron]{srinivasan2021nerv}
Pratul~P. Srinivasan, Boyang Deng, Xiuming Zhang, Matthew Tancik, Ben Mildenhall, and Jonathan~T. Barron.
\newblock {{NeRV}: Neural Reflectance and Visibility Fields for Relighting and View Synthesis}.
\newblock In \emph{Conference on Computer Vision and Pattern Recognition}, 2021.

\bibitem[Sun et~al.(2022)Sun, Sun, and Chen]{sun2022direct}
Cheng Sun, Min Sun, and Hwann-Tzong Chen.
\newblock {Direct Voxel Grid Optimization: Super-Fast Convergence for Radiance Fields Reconstruction}.
\newblock In \emph{Conference on Computer Vision and Pattern Recognition}, 2022.

\bibitem[Verbin et~al.(2022)Verbin, Hedman, Mildenhall, Zickler, Barron, and Srinivasan]{verbin2022ref}
Dor Verbin, Peter Hedman, Ben Mildenhall, Todd Zickler, Jonathan~T. Barron, and Pratul~P. Srinivasan.
\newblock {{Ref-NeRF}: Structured View-Dependent Appearance for Neural Radiance Fields}.
\newblock In \emph{Conference on Computer Vision and Pattern Recognition}, 2022.

\bibitem[Wang et~al.(2021)Wang, Liu, Liu, Theobalt, Komura, and Wang]{wang2021neus}
Peng Wang, Lingjie Liu, Yuan Liu, Christian Theobalt, Taku Komura, and Wenping Wang.
\newblock {NeuS: Learning Neural Implicit Surfaces by Volume Rendering for Multi-View Reconstruction}.
\newblock In \emph{Advances in Neural Information Processing Systems}, 2021.

\bibitem[Wood et~al.(2000)Wood, Azuma, Aldinger, Curless, Duchamp, Salesin, and Stuetzle]{wood:2000:slf}
Daniel~N. Wood, Daniel~I. Azuma, Ken Aldinger, Brian Curless, Tom Duchamp, David~H. Salesin, and Werner Stuetzle.
\newblock {Surface Light Fields for {3D} Photography}.
\newblock In \emph{ACM SIGGRAPH}, 2000.

\bibitem[Yariv et~al.(2021)Yariv, Gu, Kasten, and Lipman]{yariv2021volsdf}
Lior Yariv, Jiatao Gu, Yoni Kasten, and Yaron Lipman.
\newblock {Volume Rendering of Neural Implicit Surfaces}.
\newblock In \emph{Advances in Neural Information Processing Systems}, 2021.

\bibitem[Yariv et~al.(2023)Yariv, Hedman, Reiser, Verbin, Srinivasan, Szeliski, and Barron]{yariv-2023-bakedsdf}
Lior Yariv, Peter Hedman, Christian Reiser, Dor Verbin, Pratul~P. Srinivasan, Richard Szeliski, and Jonathan~T. Barron.
\newblock {BakedSDF: Meshing Neural SDFs for Real-Time View Synthesis}.
\newblock In \emph{ACM SIGGRAPH}, 2023.

\bibitem[Yu et~al.(2021)Yu, Li, Tancik, Li, Ng, and Kanazawa]{yu2021plenoctrees}
Alex Yu, Ruilong Li, Matthew Tancik, Hao Li, Ren Ng, and Angjoo Kanazawa.
\newblock {{PlenOctrees} For Real-Time Rendering of Neural Radiance Fields}.
\newblock In \emph{International Conference on Computer Vision}, 2021.

\bibitem[Yu et~al.(2022{\natexlab{a}})Yu, Fridovich-Keil, Tancik, Chen, Recht, and Kanazawa]{yu2021plenoxels}
Alex Yu, Sara Fridovich-Keil, Matthew Tancik, Qinhong Chen, Benjamin Recht, and Angjoo Kanazawa.
\newblock {Plenoxels: Radiance Fields Without Neural Networks}.
\newblock In \emph{Conference on Computer Vision and Pattern Recognition}, 2022{\natexlab{a}}.

\bibitem[Yu et~al.(2022{\natexlab{b}})Yu, Fridovich-Keil, Tancik, Chen, Recht, and Kanazawa]{yu_and_fridovichkeil2021plenoxels}
Alex Yu, Sara Fridovich-Keil, Matthew Tancik, Qinhong Chen, Benjamin Recht, and Angjoo Kanazawa.
\newblock {Plenoxels: Radiance Fields Without Neural Networks}.
\newblock In \emph{Conference on Computer Vision and Pattern Recognition}, 2022{\natexlab{b}}.

\bibitem[Zhang et~al.(2021)Zhang, Luan, Wang, Bala, and Snavely]{zhang2021physg}
Kai Zhang, Fujun Luan, Qianqian Wang, Kavita Bala, and Noah Snavely.
\newblock {{PhySG}: Inverse Rendering with Spherical Gaussians for Physics-Based Material Editing and Relighting}.
\newblock In \emph{Conference on Computer Vision and Pattern Recognition}, 2021.

\bibitem[Zhang et~al.(2018)Zhang, Isola, Efros, Shechtman, and Wang]{zhang2018lpips}
Richard Zhang, Phillip Isola, Alexei~A. Efros, Eli Shechtman, and Oliver Wang.
\newblock {The Unreasonable Effectiveness of Deep Features as a Perceptual Metric}.
\newblock In \emph{Conference on Computer Vision and Pattern Recognition}, 2018.

\end{thebibliography}
}

\clearpage
\setcounter{page}{1}
\maketitlesupplementary

In this supplementary material, we provide the following elements:
\begin{itemize}
    \item Details about the parameterisation of the bound gaussians optimized during our joint refinement strategy.
    \item Additional implementation details.
    \item Detailed quantitative results for real-time rendering of real scenes, and mesh rendering ablation.
\end{itemize}
We also provide a video that offers an overview of the approach and showcases additional qualitative results. Specifically, the video demonstrates how SuGaR meshes can be used to animate Gaussian Splatting representations.  

\section{Parameterisation of Gaussians bound to the surface}
\label{sec:suppmat_bound_gaussians_param}

As we explained in Section~\ref{sec:method}, once we have extracted the mesh from the Gaussian Splatting representation, we refine this mesh by binding new Gaussians to the mesh triangles and optimize the Gaussians and the mesh jointly using the Gaussian Splatting rasterizer. 
To keep the Gaussians flat and aligned with the mesh triangles, we explicitly compute the means of the Gaussians from the mesh vertices using predefined barycentric coordinates in the corresponding triangles during optimization. 
Also, the Gaussians have only 2 learnable scaling factors instead of 3 and only 1 learnable 2D rotation. Indeed, we do not optimize a full quaternion that would encode a 3D rotation, as performed in~\cite{kerbl3Dgaussians}; Instead, we optimize a 2D rotation in the plane of the triangle. Therefore, the Gaussians stay aligned with the mesh triangles, but are allowed to rotate on the local surface. Like the original model, we also optimize an opacity value and a set of spherical harmonics for every Gaussian to encode the color emitted in all directions.

In practice, for each Gaussian, we optimize a learnable complex number $x+iy$ rather than a quaternion, encoding the 2D rotation inside the triangle's plane. During optimization, we still need to compute an explicit 3D quaternion encoding the 3D rotation of the Gaussians in the \textit{world space} to apply the rasterizer. 
To recover the full 3D quaternion, we proceed as follows: For any 3D Gaussian $g$, we first compute the matrix $R=[R^{(0)}, R^{(1)}, R^{(2)}] \in \IR^{3\times 3}$ encoding the rotation of its corresponding triangle: We select as the first column $R^{(0)}$ of the matrix the normal of the triangle, and as the second column $R^{(1)}$ a fixed edge of the triangle. We compute the third column $R^{(2)}$ with a cross-product.
Then, we compute the matrix $R_g$ encoding the full 3D rotation of the Gaussian by applying the learned 2D complex number to the rotation of the triangle, as follows: $R_g^{(0)} = R^{(0)}, R_g^{(1)}=x'R^{(1)} + y'R^{(2)}$ and $R_g^{(2)} = -y'R^{(1)} + x'R^{(2)}$, where $x' = \frac{x}{|x^2 + y^2|}$ and $y' = \frac{y}{|x^2 + y^2|}$.

\paragraph{Adjusting parameters for edition.} Because our learned complex numbers represent rotations in the space of the corresponding triangles, our representation is robust to mesh edition or animation: When editing the underlying mesh at inference, there is no need to update the learned 2D rotations as they remain the same when rotating or moving triangles.

Conversely, when scaling or deforming a mesh, the triangle sizes might change, necessitating adjustments to the learned scaling factors of the bound surface Gaussians. For example, if the mesh size doubles, all Gaussian scaling factors should similarly be multiplied by 2. 
In our implementation, when editing the mesh, we modify in real-time the learned scaling factors of a bound surface Gaussian by multiplying them by the ratio between (a) the average length of the triangle's sides after modification and (b) the average length of the original triangle's sides.

\section{Additional implementation details}

\paragraph{Implementation} We implemented our model with PyTorch~\cite{paszke-nips19-pytorch} and use 3D data processing tools from PyTorch3D~\cite{ravi-arxiv20-accelerating3ddeeplearning}. We also use the differentiable Gaussian Splatting rasterizer from the original 3D Gaussian Splatting paper~\cite{kerbl3Dgaussians}. We thank the authors for providing this amazing tool.

\paragraph{Mesh extraction.} In practice, we apply two Poisson reconstructions for mesh extraction: one for foreground points, and one for background points. 
We define foreground points as points located inside the bounding box of all training camera poses, and background points as points located outside.  
We chose this simple distinction between foreground and background in order to design an approach as general as possible. 
However, depending on the content of the scene and the main objects to reconstruct, defining a custom bounding box for foreground points could improve the quality and precision of the extracted mesh.

\paragraph{Joint refinement.} During joint refinement, we also compute a normal consistency term on the mesh's faces to further regularize the surface. This term doesn't affect performance in terms of PSNR, SSIM, or LPIPS. However, it does marginally enhance visual quality by promoting smoother surfaces.

\begin{table}
  \centering
  {\small
  \begin{tabular}{@{}lcccc@{}}
    \toprule
      %
     %
      & PSNR $\uparrow$ & SSIM $\uparrow$ & LPIPS $\downarrow$ \\
    \midrule
    Plenoxels~\cite{yu_and_fridovichkeil2021plenoxels} & 21.07 & 0.719 & 0.379 \\
    INGP-Base~\cite{mueller2022instantngp} & 21.72 & 0.723 & 0.330 \\
    INGP-Big~\cite{mueller2022instantngp} & \cellcolor{yellow!25}21.92 & 0.744 & 0.304 \\
    Mip-NeRF360~\cite{barron2022mipnerf360} & \cellcolor{orange!25}22.22 & 0.758 & 0.257 \\
    3DGS~\cite{kerbl3Dgaussians} & \cellcolor{red!25}\textbf{23.14} & \cellcolor{red!25}\textbf{0.841} & \cellcolor{red!25}\textbf{0.183} \\
    \midrule
    R-SuGaR-2K (Ours) & 19.70 & 0.743 & 0.284 \\
    R-SuGaR-7K (Ours) & 21.09 & \cellcolor{yellow!25}0.786 & \cellcolor{yellow!25}0.233 \\
    R-SuGaR-15K (Ours) & 21.58 & \cellcolor{orange!25}0.795 & \cellcolor{orange!25}0.219 \\
    \bottomrule
  \end{tabular}
  }
  \caption{\textbf{Quantitative evaluation on Tanks\&Temples~\cite{knapitsch-2017-tanksandtemples}.} SuGaR is not as good as as vanilla 3D Gaussian Splatting in terms of rendering quality as it relies on a mesh but higher than the other methods that do not recover a mesh.}
  \label{tab:nvsmetrics_tandt}
\end{table}
\begin{table}
  \centering
  {\small
  \begin{tabular}{@{}lcccc@{}}
    \toprule
      %
     %
      & PSNR $\uparrow$ & SSIM $\uparrow$ & LPIPS $\downarrow$\\
    \midrule
    Plenoxels~\cite{yu_and_fridovichkeil2021plenoxels} & 23.06 & 0.794 & 0.510 \\
    INGP-Base~\cite{mueller2022instantngp} & 23.62 & 0.796 & 0.423 \\
    INGP-Big~\cite{mueller2022instantngp} & 24.96 & 0.817 & 0.390 \\
    Mip-NeRF360~\cite{barron2022mipnerf360} & \cellcolor{orange!25}29.40 & \cellcolor{orange!25}0.901 & \cellcolor{orange!25}0.244 \\
    3DGS~\cite{kerbl3Dgaussians} & \cellcolor{red!25}\textbf{29.41} & \cellcolor{red!25}\textbf{0.903} & \cellcolor{red!25}\textbf{0.242} \\
    \midrule
    R-SuGaR-2K (Ours) & 27.31 & 0.873 & 0.303 \\
    R-SuGaR-7K (Ours) & \cellcolor{yellow!25}29.30 & \cellcolor{yellow!25}0.893 & 0.273 \\
    R-SuGaR-15K (Ours) & \cellcolor{red!25}\textbf{29.41} & \cellcolor{yellow!25}0.893 & \cellcolor{yellow!25}0.267 \\
    \bottomrule
  \end{tabular}
  }
  \caption{\textbf{Quantitative evaluation on DeepBlending~\cite{hedman-2018-deepblending}.} SuGaR is not as good as as vanilla 3D Gaussian Splatting in terms of rendering quality as it relies on a mesh but higher than the other methods that do not recover a mesh.}
  \label{tab:nvsmetrics_db}
\end{table}

\section{Additional Results for Real-Time Rendering of Real Scenes}
\label{suppmat_realtimerendering}

We compute the standard metrics PSNR, SSIM and LPIPS~\cite{zhang2018lpips} to evaluate the quality of SuGaR's rendering using our extracted meshes and their bound surface Gaussians. 
Results on the Mip-NeRF360 dataset are given in Table~\ref{tab:nvsmetrics_mipnerf360} in the main paper. Results on Tanks\&Temple and DeepBlending are given in Tables~\ref{tab:nvsmetrics_tandt} and ~\ref{tab:nvsmetrics_db}.
Tables~\ref{tab:nvsmetrics_psnr},~\ref{tab:nvsmetrics_ssim} and ~\ref{tab:nvsmetrics_lpips} provide the detailed results for all scenes in the datasets.

\section{Additional Results for Mesh Renderig Ablation}
\label{suppmat_renderingablation}

We provide additional qualitative results to illustrate how various parameters impact rendering performance.

\begin{figure}
  \centering
  \begin{subfigure}{0.32\linewidth}
        \includegraphics[width=\linewidth]{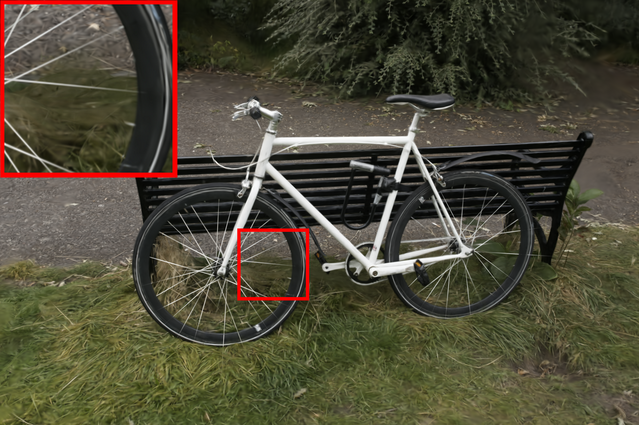}
        \caption{2,000 iterations}
    \end{subfigure}
    \hfill
   \begin{subfigure}{0.32\linewidth}
        \includegraphics[width=\linewidth]{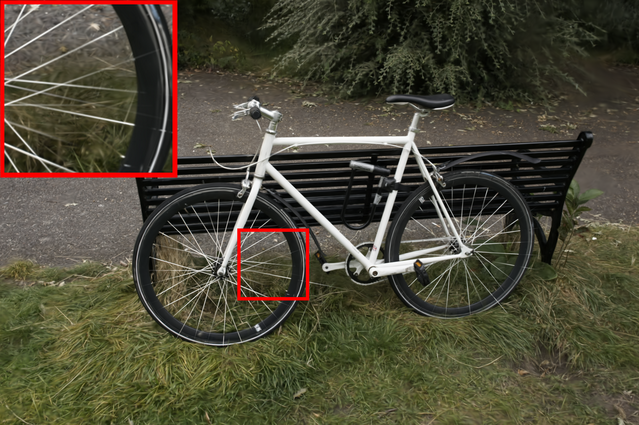}
        \caption{7,000 iterations}
    \end{subfigure}
    \hfill
    \begin{subfigure}{0.32\linewidth}
        \includegraphics[width=\linewidth]{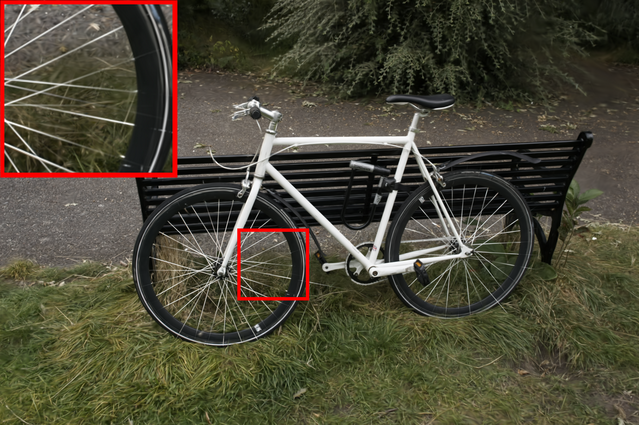}
        \caption{15,000 iterations}
    \end{subfigure}
   \caption{\textbf{Refined SuGaR renderings with different numbers of refinement iterations.} 2,000 iterations are usually enough to obtain high quality rendering \textbf{(a)}, since the extracted mesh ``textured'' with surface Gaussians is already an excellent initialization for optimizing the model. However, further refinement helps the Gaussians to capture texturing details and reconstruct extremely thin geometry that is finer that the resolution of the mesh, such as the spokes of the bicycle, as seen in \textbf{(b)}, \textbf{(c)}.}
   \label{fig:refinement}
\end{figure}
\begin{figure}[t]
  \centering
  \begin{subfigure}{0.32\linewidth}
        \includegraphics[width=\linewidth]{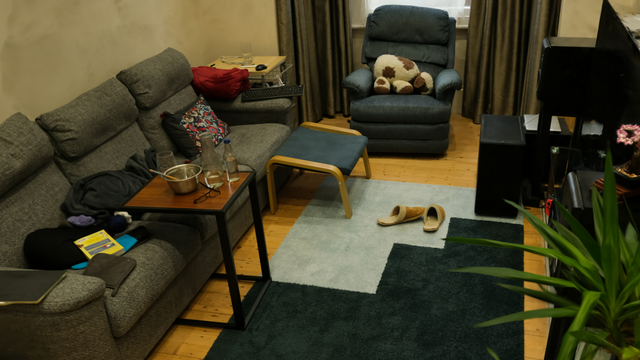}%
        \vspace{.1em}
        \includegraphics[width=\linewidth]{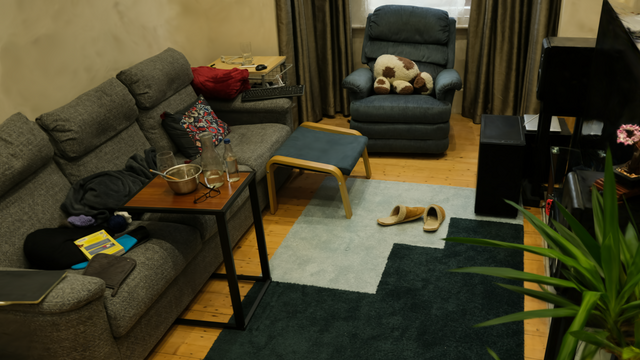}%
        \vspace{.1em}
        \caption{SuGaR render}
    \end{subfigure}
    \hfill
      \begin{subfigure}{0.32\linewidth}
        \includegraphics[width=\linewidth]{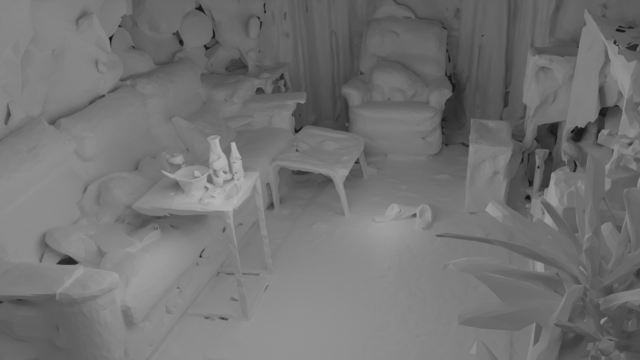}%
        \vspace{.1em}
        \includegraphics[width=\linewidth]{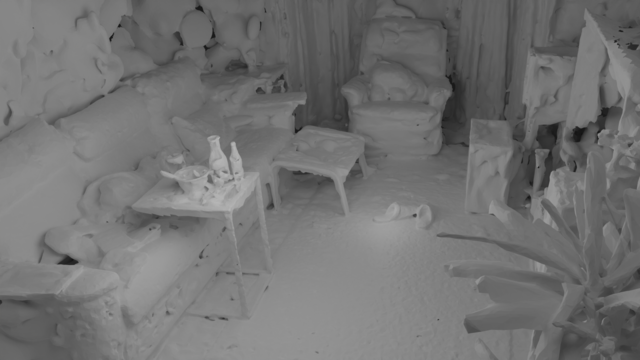}%
        \vspace{.1em}
        \caption{Mesh (No texture)}
    \end{subfigure}
    \hfill
   \begin{subfigure}{0.32\linewidth}
        \includegraphics[width=\linewidth]{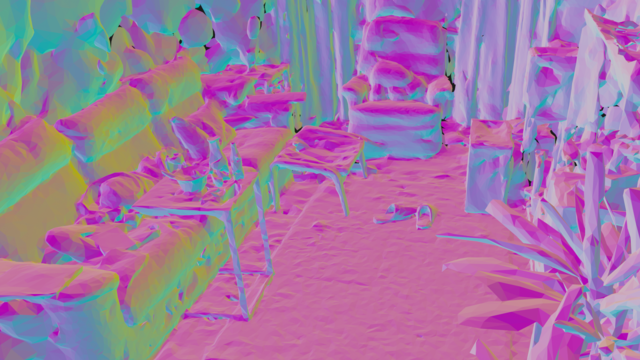}%
        \vspace{.1em}
        \includegraphics[width=\linewidth]{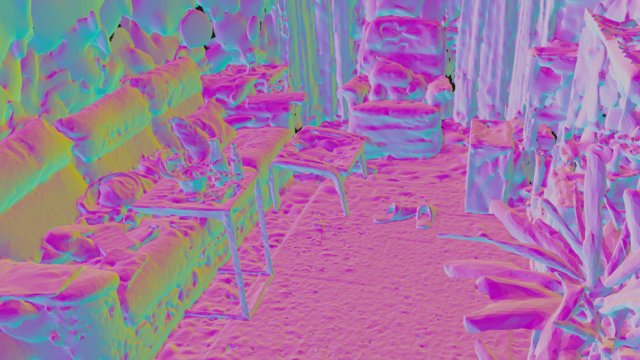}%
        \vspace{.1em}
        \caption{Mesh normals}
    \end{subfigure}
   \caption{\textbf{SuGaR renderings} with (\textbf{top:}) 200,000 and (\textbf{bottom:}) 1,000,000 vertices. Even with low-poly meshes, the 3D Gaussians bound to the mesh produce high quality renderings. Moreover, low-poly meshes help to better regularize the surface.}
   \label{fig:render_number_of_triangles}
\end{figure}
\begin{figure}[t]
  \centering
   \begin{subfigure}{0.24\linewidth}
        \includegraphics[height=0.7\linewidth,width=\linewidth]{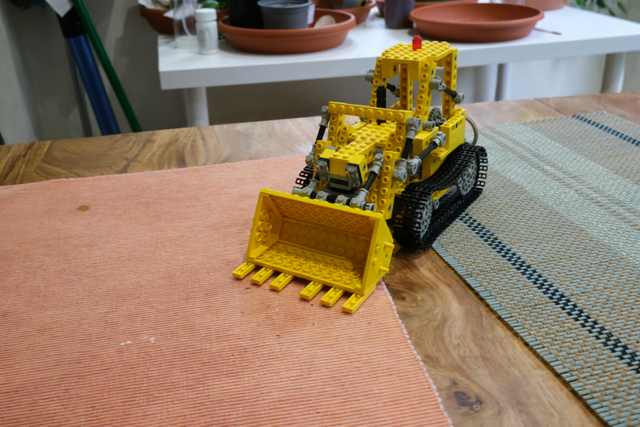}%
        \vspace{.1em}
        \includegraphics[height=0.7\linewidth,width=\linewidth]{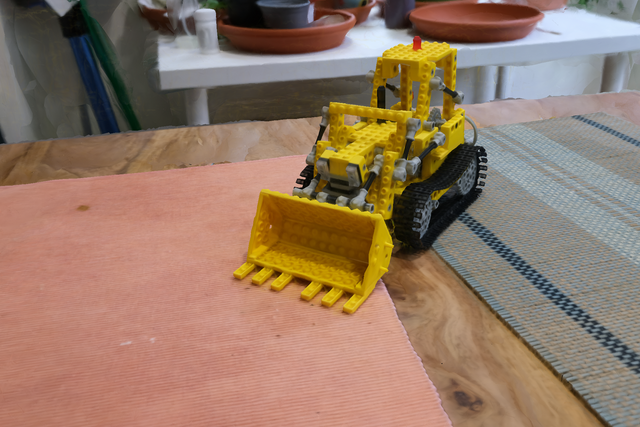}%
        \vspace{.1em}
    \end{subfigure}
    \hfill
   \begin{subfigure}{0.24\linewidth}
        \includegraphics[height=0.7\linewidth,width=\linewidth]{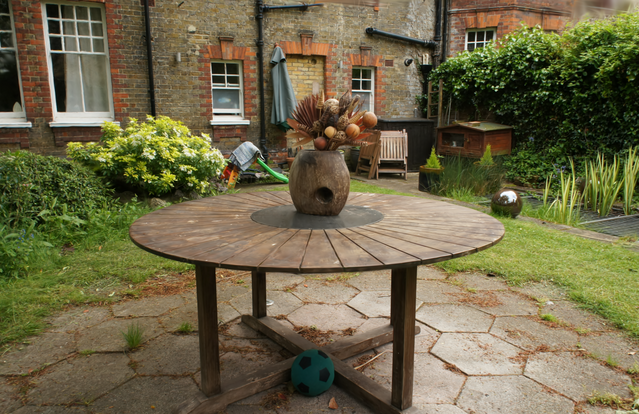}%
        \vspace{.1em}
        \includegraphics[height=0.7\linewidth,width=\linewidth]{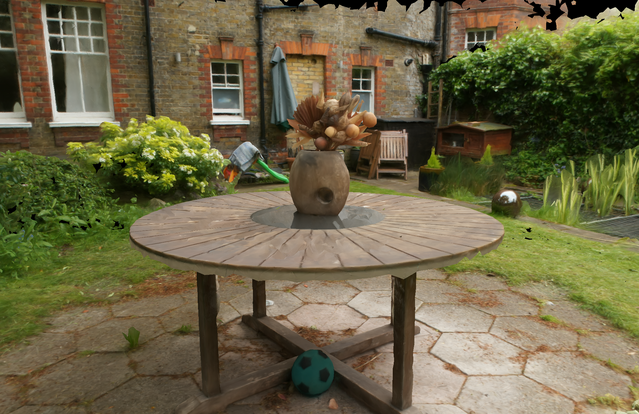}%
        \vspace{.1em}
    \end{subfigure}
    \hfill
   \begin{subfigure}{0.24\linewidth}
        \includegraphics[height=0.7\linewidth,width=\linewidth]{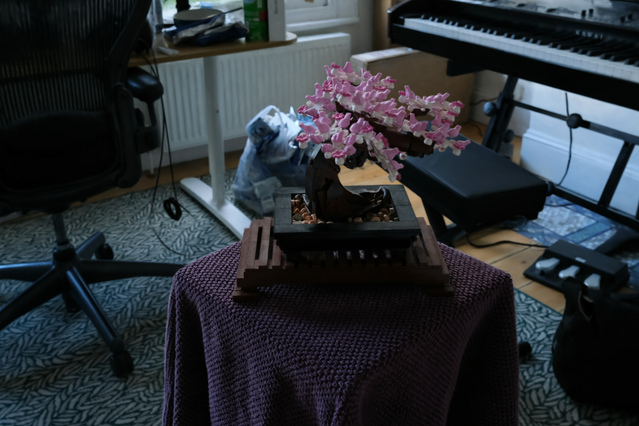}%
        \vspace{.1em}
        \includegraphics[height=0.7\linewidth,width=\linewidth]{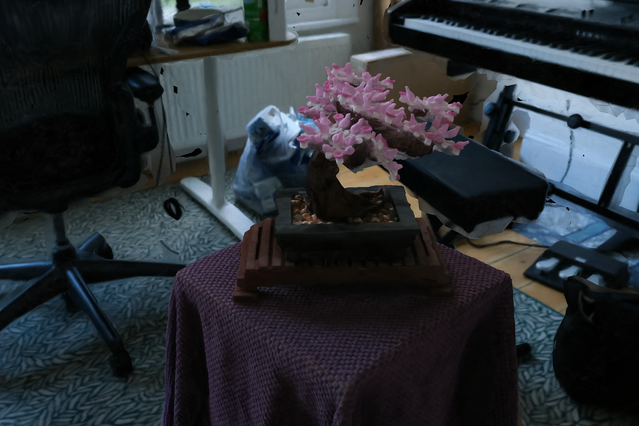}%
        \vspace{.1em}
    \end{subfigure}
    \hfill
   \begin{subfigure}{0.24\linewidth}
        \includegraphics[height=0.7\linewidth,width=\linewidth]{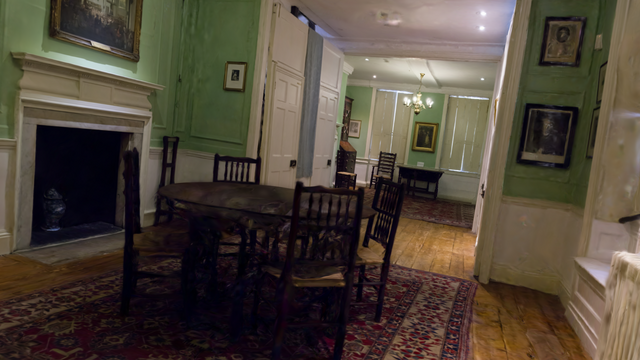}%
        \vspace{.1em}
        \includegraphics[height=0.7\linewidth,width=\linewidth]{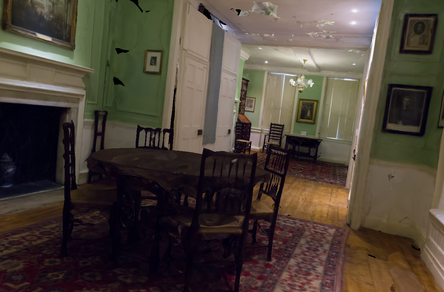}%
        \vspace{.1em}
    \end{subfigure}
   \caption{\textbf{Qualitative comparison between (top:) a traditional UV texture optimized from training images, and (bottom:) the bound surface Gaussians.} Even though high resolution UV textures have good quality and can be rendered with our meshes using any traditional software, using 3D Gaussians bound to the surface of the mesh greatly improves the rendering quality. Meshes in these images have 200,000 vertices only.}
   \label{fig:traditional_texture}
\end{figure}

First, we provide in Figure~\ref{fig:refinement} a simple example showing how the Gaussians constrained to remain on the surface during refinement greatly increase the rendering quality as they play the role of an efficient texturing tool and help reconstructing very fine details missing in the extracted mesh.

Then, in Figure~\ref{fig:render_number_of_triangles} {}we illustrate how the resolution of the mesh extraction, i.e., the number of triangles, modifies the rendering quality. For fair comparison, we increase the number of surface-aligned Gaussians per triangle when we decrease the number of triangles. Results show that increasing the number of vertices increases the quality of rendering with surface Gaussians, but meshes with lower triangles are already able to reach state of the art results.

Finally, Figure~\ref{fig:traditional_texture} illustrates the benefits of using Gaussians aligned on the surface as a texturing tool for rendering meshes. To this end, we also optimize traditional UV textures on our meshes using differentiable mesh rendering with traditional triangle rasterization. Even though rendering with surface-aligned Gaussians provides better performance, rendering our meshes with traditional UV textures still produces satisfying results, which further illustrates the quality of our extracted meshes. 

\begin{table*}
  \centering
  {\footnotesize 
  \begin{tabular}{@{}lcccccccccccc@{}}
    \toprule
      %
     %
     \multicolumn{1}{c}{} & \multicolumn{7}{c}{Mip-NeRF360~\cite{barron2022mipnerf360}} & \multicolumn{2}{c}{DeepBlending~\cite{hedman-2018-deepblending}} & \multicolumn{2}{c}{Tanks\&Temples~\cite{knapitsch-2017-tanksandtemples}} \\
     \cmidrule(r){2-8} \cmidrule(r){9-10} \cmidrule(r){11-12}
        & Garden & Kitchen & Room & Bicycle & Counter & Bonsai & Stump & Playroom & Dr. Johnson & Train & Truck \\ 
        \midrule
        \multicolumn{10}{l}{\textbf{200K vertices}} \\
        \midrule
        R-SuGaR-2K & 23.30 & 25.74 & 27.58 & 21.53 & 24.41 & 26.50 & 23.45 & 27.83 & 26.51 & 18.15 & 21.03 \\
        R-SuGaR-7K & 24.99 & 28.78 & 29.47 & 22.69 & 26.86 & 29.33 & \cellcolor{yellow!25}24.45 & 30.02 & 28.41 & 19.82 & 22.31 \\
        R-SuGaR-15K & \cellcolor{orange!25}25.29 & \cellcolor{orange!25}29.38 & \cellcolor{orange!25}29.95 & \cellcolor{orange!25}22.91 & \cellcolor{orange!25}27.47 & \cellcolor{orange!25}30.42 & \cellcolor{orange!25}24.55 & \cellcolor{yellow!25}30.08 & \cellcolor{orange!25}28.59 & \cellcolor{orange!25}20.40 & \cellcolor{orange!25}22.65 \\
        \midrule
        \multicolumn{10}{l}{\textbf{1M vertices}} \\
        \midrule
        R-SuGaR-2K & 23.56 & 26.15 & 27.68 & 21.80 & 24.62 & 26.70 & 23.56 & 27.93 & 26.70 & 18.32 & 21.09 \\
        R-SuGaR-7K & \cellcolor{yellow!25}25.06 & \cellcolor{yellow!25}28.96 & \cellcolor{yellow!25}29.57 & \cellcolor{yellow!25}22.86 & \cellcolor{yellow!25}26.92 & \cellcolor{yellow!25}29.47 & \cellcolor{orange!25}24.55 & \cellcolor{red!25}\textbf{30.13} & \cellcolor{yellow!25}28.47 & \cellcolor{yellow!25}19.85 & \cellcolor{yellow!25}22.34 \\
        R-SuGaR-15K & \cellcolor{red!25}\textbf{25.36} & \cellcolor{red!25}\textbf{29.56} & \cellcolor{red!25}\textbf{30.03} & \cellcolor{red!25}\textbf{23.14} & \cellcolor{red!25}\textbf{27.62} & \cellcolor{red!25}\textbf{30.51} & \cellcolor{red!25}\textbf{24.70} & \cellcolor{orange!25}30.12 & \cellcolor{red!25}\textbf{28.71} & \cellcolor{red!25}\textbf{20.50} & \cellcolor{red!25}\textbf{22.67} \\
        \bottomrule
  \end{tabular}
  }
  \caption{\textbf{Quantitative evaluation of rendering quality in terms of PSNR on all scenes.} A higher PSNR indicates better rendering quality. We adjust the number of bound surface-aligned Gaussians per triangle when we reduce the number of vertices, aiming for a similar count across all models. Results show that increasing the number of vertices (\ie increasing the resolution of the geometry) increases the quality of rendering with surface Gaussians, but meshes with less triangles are already able to reach state of the art results.}
  \label{tab:nvsmetrics_psnr}
\end{table*}
\begin{table*}
  \centering
  {\footnotesize 
  \begin{tabular}{@{}lcccccccccccc@{}}
    \toprule
      %
     %
     \multicolumn{1}{c}{} & \multicolumn{7}{c}{Mip-NeRF360~\cite{barron2022mipnerf360}} & \multicolumn{2}{c}{DeepBlending~\cite{hedman-2018-deepblending}} & \multicolumn{2}{c}{Tanks\&Temples~\cite{knapitsch-2017-tanksandtemples}} \\
     \cmidrule(r){2-8} \cmidrule(r){9-10} \cmidrule(r){11-12}
        & Garden & Kitchen & Room & Bicycle & Counter & Bonsai & Stump & Playroom & Dr. Johnson & Train & Truck \\ 
        \midrule
        \multicolumn{10}{l}{\textbf{200K vertices}} \\
        \midrule
        R-SuGaR-2K & 0.713 & 0.859 & 0.881 & 0.572 & 0.844 & 0.895 & 0.641 & 0.883 & 0.864 & 0.694 & 0.787 \\
        R-SuGaR-7K & 0.762 & 0.901 & \cellcolor{yellow!25}0.904 & 0.621 & 0.883 & \cellcolor{yellow!25}0.926 & 0.679 & \cellcolor{orange!25}0.898 & \cellcolor{orange!25}0.888 & 0.749 & \cellcolor{orange!25}0.822 \\
        R-SuGaR-15K & \cellcolor{orange!25}0.771 & \cellcolor{orange!25}0.907 & \cellcolor{red!25}\textbf{0.909} & \cellcolor{orange!25}0.631 & \cellcolor{orange!25}0.890 & \cellcolor{red!25}\textbf{0.933} & \cellcolor{orange!25}0.681 & \cellcolor{yellow!25}0.897 & \cellcolor{orange!25}0.888 & \cellcolor{orange!25}0.763 & \cellcolor{red!25}\textbf{0.827} \\
        \midrule
        \multicolumn{10}{l}{\textbf{1M vertices}} \\
        \midrule
        R-SuGaR-2K & 0.719 & 0.866 & 0.882 & 0.583 & 0.846 & 0.894 & 0.642 & 0.883 & 0.863 & 0.698 & 0.788 \\
        R-SuGaR-7K & \cellcolor{yellow!25}0.764 & \cellcolor{yellow!25}0.903 & \cellcolor{orange!25}0.905 & \cellcolor{yellow!25}0.628 & \cellcolor{yellow!25}0.884 & 0.925 & \cellcolor{yellow!25}0.680 & \cellcolor{red!25}\textbf{0.899} & \cellcolor{yellow!25}0.887 & \cellcolor{yellow!25}0.750 & \cellcolor{yellow!25}0.821 \\
        R-SuGaR-15K & \cellcolor{red!25}\textbf{0.775} & \cellcolor{red!25}\textbf{0.908} & \cellcolor{red!25}\textbf{0.909} & \cellcolor{red!25}\textbf{0.640} & \cellcolor{red!25}\textbf{0.891} & \cellcolor{orange!25}0.932 & \cellcolor{red!25}\textbf{0.683} & \cellcolor{orange!25}0.898 & \cellcolor{red!25}\textbf{0.889} & \cellcolor{red!25}\textbf{0.764} & \cellcolor{red!25}\textbf{0.827} \\
        \bottomrule
  \end{tabular}
  }
  \caption{\textbf{Quantitative evaluation of rendering quality in terms of SSIM on all scenes.} A higher SSIM indicates better rendering quality. We adjust the number of bound surface-aligned Gaussians per triangle when we reduce the number of vertices, aiming for a similar count across all models. Results show that increasing the number of vertices (\ie increasing the resolution of the geometry) increases the quality of rendering with surface Gaussians, but meshes with less triangles are already able to reach state of the art results.}
  \label{tab:nvsmetrics_ssim}
\end{table*}
\begin{table*}
  \centering
  {\footnotesize 
  \begin{tabular}{@{}lcccccccccccc@{}}
    \toprule
      %
     %
     \multicolumn{1}{c}{} & \multicolumn{7}{c}{Mip-NeRF360~\cite{barron2022mipnerf360}} & \multicolumn{2}{c}{DeepBlending~\cite{hedman-2018-deepblending}} & \multicolumn{2}{c}{Tanks\&Temples~\cite{knapitsch-2017-tanksandtemples}} \\
     \cmidrule(r){2-8} \cmidrule(r){9-10} \cmidrule(r){11-12}
        & Garden & Kitchen & Room & Bicycle & Counter & Bonsai & Stump & Playroom & Dr. Johnson & Train & Truck \\ 
        \midrule
        \multicolumn{10}{l}{\textbf{200K vertices}} \\
        \midrule
        R-SuGaR-2K & 0.280 & 0.221 & 0.280 & 0.413 & 0.288 & 0.259 & 0.390 & 0.284 & 0.314 & 0.335 & 0.235 \\
        R-SuGaR-7K & \cellcolor{yellow!25}0.232 & 0.175 & \cellcolor{yellow!25}0.252 & 0.363 & \cellcolor{orange!25}0.245 & \cellcolor{yellow!25}0.228 & \cellcolor{yellow!25}0.345 & \cellcolor{orange!25}0.260 & \cellcolor{yellow!25}0.277 & \cellcolor{yellow!25}0.274 & \cellcolor{yellow!25}0.187 \\
        R-SuGaR-15K & \cellcolor{red!25}\textbf{0.218} & \cellcolor{orange!25}0.166 & \cellcolor{red!25}\textbf{0.243} & \cellcolor{orange!25}0.349 & \cellcolor{red!25}\textbf{0.234} & \cellcolor{red!25}\textbf{0.219} & \cellcolor{red!25}\textbf{0.336} & \cellcolor{red!25}\textbf{0.257} & \cellcolor{red!25}\textbf{0.268} & \cellcolor{red!25}\textbf{0.258} & \cellcolor{red!25}\textbf{0.174} \\
        \midrule
        \multicolumn{10}{l}{\textbf{1M vertices}} \\
        \midrule
        R-SuGaR-2K & 0.281 & 0.215 & 0.282 & 0.408 & \cellcolor{yellow!25}0.287 & 0.262 & 0.391 & 0.286 & 0.319 & 0.333 & 0.236 \\
        R-SuGaR-7K & 0.233 & \cellcolor{yellow!25}0.173 & 0.253 & \cellcolor{yellow!25}0.360 & \cellcolor{orange!25}0.245 & 0.231 & 0.347 & 0.265 & 0.282 & 0.275 & 0.190 \\
        R-SuGaR-15K & \cellcolor{orange!25}0.220 & \cellcolor{red!25}\textbf{0.165} & \cellcolor{orange!25}0.246 & \cellcolor{red!25}\textbf{0.345} & \cellcolor{red!25}\textbf{0.234} & \cellcolor{orange!25}0.221 & \cellcolor{orange!25}0.338 & \cellcolor{yellow!25}0.261 & \cellcolor{orange!25}0.273 & \cellcolor{orange!25}0.260 & \cellcolor{orange!25}0.178 \\
        \bottomrule
  \end{tabular}
  }
  \caption{\textbf{Quantitative evaluation of rendering quality in terms of LPIPS~\cite{zhang2018lpips} on all scenes.} A lower LPIPS indicates better rendering quality. We adjust the number of bound surface-aligned Gaussians per triangle when we reduce the number of vertices, aiming for a similar count across all models. The results indicate that the stronger regularity due to a smaller number of vertices leads to smoother surfaces and higher LPIPS metrics when using the bound Gaussians.
  }
  \label{tab:nvsmetrics_lpips}
\end{table*}

\end{document}